\documentclass[12pt,a4paper,twocolumn,aps,fleqn,nofootinbib,superscriptaddress,longbibliography]{revtex4}
\usepackage[utf8]{inputenc}
\usepackage{epsf}
\usepackage{latexsym,amssymb,euscript}
\usepackage[dvips]{graphicx}
\usepackage[fleqn]{amsmath}
\usepackage{nicefrac}
\usepackage{slashed}
\usepackage{booktabs}
\usepackage{hyperref}
\usepackage{braket}
\usepackage{chngcntr}
\usepackage{bbold}
\usepackage{graphics}
\usepackage{graphicx}
\usepackage{mciteplus}
\usepackage{pdfpages}
\usepackage{setspace}
\usepackage[titletoc]{appendix}
\graphicspath{{./figures/}}
\hypersetup{
 linktocpage = true,
 urlcolor = purple,
 colorlinks = true,
 linkcolor = purple,
 anchorcolor = purple,
 citecolor = purple,
 pdfstartview = {XYZ null null 1.25} 
           }
\usepackage[left=1.5cm, right=1.5cm, top=3.0cm, bottom=2.5cm]{geometry}
\usepackage{pstricks}
\usepackage{color}
\usepackage{float}
\usepackage{academicons}
\definecolor{orcidlogocol}{HTML}{A6CE39}
\usepackage{fancyhdr}
\newcommand{\drv}{{\rm d}}

\newcommand{\as}{\alpha_s}
\newcommand{\LQCD}{\Lambda_{\rm QCD}}
\newcommand{\MSb}{\overline{\rm MS}}

\newcommand{\DY}{\Delta Y}

\newcommand{\tild}{~}
\newcommand{\tcite}[1]{~\cite{#1}}
\newcommand{\tref}[1]{~\ref{#1}}
\newcommand{\eref}[1]{~\eqref{#1}}

\newcommand{\tarr}{
\begin{array}}
\newcommand{\earr}{\end{array}}

\newcommand{\revdone}[1]{~\textcolor{black}{#1}}

\newcommand{\orcidFGC}{\href{https://orcid.org/0000-0003-3299-2203}{\includegraphics[scale=0.1]{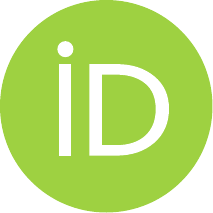}}}

\newcommand{\orcidMF}{\href{https://orcid.org/0000-0002-2408-2210}{\includegraphics[scale=0.1]{figures/logo-orcid.pdf}}}

\newcommand{\orcidDYI}{\href{https://orcid.org/0000-0001-5701-4364}{\includegraphics[scale=0.1]{figures/logo-orcid.pdf}}}

\newcommand{\orcidMMAM}{\href{http://orcid.org/0000-0002-6449-2257}{\includegraphics[scale=0.1]{figures/logo-orcid.pdf}}}

\newcommand{\orcidAP}{\href{https://orcid.org/0000-0001-8984-3036}{\includegraphics[scale=0.1]{figures/logo-orcid.pdf}}}

\begin{document}


\title{
  {\Large \bf Bottom-flavored inclusive emissions \\ in the variable-flavor number scheme: a high-energy analysis}
}

\author{Francesco Giovanni Celiberto \orcidFGC}
\email{fceliberto@ectstar.eu}
\affiliation{\setstretch{1.0}European Centre for Theoretical Studies in Nuclear Physics and Related Areas (ECT*), I-38123 Villazzano, Trento, Italy}
\affiliation{\setstretch{1.0}Fondazione Bruno Kessler (FBK), I-38123 Povo, Trento, Italy}
\affiliation{\setstretch{1.0}INFN-TIFPA Trento Institute of Fundamental Physics and Applications, I-38123 Povo, Trento, Italy}

\author{Michael Fucilla \orcidMF}
\email{michael.fucilla@unical.it}
\affiliation{\setstretch{1.0}Dipartimento di Fisica, Universit\`a della Calabria, I-87036 Arcavacata di Rende, Cosenza, Italy}
\affiliation{\setstretch{1.0}Istituto Nazionale di Fisica Nucleare, Gruppo collegato di Cosenza I-87036 Arcavacata di Rende, Cosenza, Italy}

\author{\\Dmitry Yu. Ivanov \orcidDYI}
\email{d-ivanov@math.nsc.ru}
\affiliation{\setstretch{1.0}Sobolev Institute of Mathematics, 630090 Novosibirsk, Russia}

\author{Mohammed M.A. Mohammed \orcidMMAM}
\email{mohammed.maher@unical.it}
\affiliation{\setstretch{1.0}Dipartimento di Fisica, Universit\`a della Calabria, I-87036 Arcavacata di Rende, Cosenza, Italy}
\affiliation{\setstretch{1.0}Istituto Nazionale di Fisica Nucleare, Gruppo collegato di Cosenza I-87036 Arcavacata di Rende, Cosenza, Italy}

\author{Alessandro Papa \orcidAP}
\email{alessandro.papa@fis.unical.it}
\affiliation{\setstretch{1.0}Dipartimento di Fisica, Universit\`a della Calabria, I-87036 Arcavacata di Rende, Cosenza, Italy}
\affiliation{\setstretch{1.0}Istituto Nazionale di Fisica Nucleare, Gruppo collegato di Cosenza I-87036 Arcavacata di Rende, Cosenza, Italy}

\begin{abstract}
\vspace{0.75cm}
\begin{center}
 {\bf Abstract}
\end{center}
\vspace{0.75cm}
\hrule \vspace{0.50cm}
\footnotesize \setstretch{1.0}
We propose the inclusive semi-hard production, in proton-proton collisions, of two bottom-flavored hadrons, as well as of a single bottom-flavored hadron accompanied by a light jet, as novel channels for the manifestation of stabilization effects of the high-energy resummation under next-to-leading-order corrections. Our formalism relies on a hybrid high-energy and collinear factorization, where the BFKL resummation of leading and next-to-leading energy logarithms is used together with collinear factorization.
We present results for cross sections and azimuthal correlations differential in rapidity, which are widely recognized as standard observables where to hunt for distinctive signals of the BFKL dynamics. We propose the study of double differential distributions in the transverse momenta of final-state particles as a common basis to investigate the interplay of different kinds of resummation mechanisms.
\vspace{0.50cm} \hrule
\vspace{0.75cm}
{
 \setlength{\parindent}{0pt}
}
\end{abstract}


\maketitle

\newpage

\begingroup
 \hypersetup{linktoc = page, 
             }
 \phantom{.}\\\phantom{.}\\\phantom{.}
 { 
 \tableofcontents
 }
 \phantom{.}\\\phantom{.}\\\phantom{.}
\endgroup


\section{Introductory remarks}
\label{sec:intro}
Heavy-flavor physics is unanimously recognized as one of the most fertile grounds where to investigate modern particle physics. On one hand, heavy-flavor experiments are relevant in the search for New Physics beyond the Standard Model (BSM), where heavy quarks are expected to be produced in association with BSM particles. On the other hand, charm and bottom quarks place themselves in a region where perturbative Quantum Chromodynamics (QCD) is applicable and their production channels can be used to perform precise tests of strong interactions.

Special attention is deserved by the production in hadronic collisions of the heaviest quark species that can hadronize, the bottom one. The standard collinear description of the $b \bar{b}$ production at next-to-leading order (NLO) was set up long time ago\tcite{Nason:1987xz,Nason:1989zy,Beenakker:1988bq}, but only recently fully differential distributions were investigated with next-to-NLO perturbative accuracy\tcite{Catani:2020kkl}. 
The study of bottom-flavor phenomenology in terms of QCD precision calculations is a quite challenging research activity. Here, the value of the $b$-quark mass, $m_b$, plays a key role. It essentially represents a threshold that determines the transition region between two different schemes. 
At low transverse-momentum values of the observed bottom-flavored object (a hadron or a heavy jet), namely when $|\vec{p}_T| \lesssim m_b$, a description in terms of the so-called fixed-flavor number scheme (FFNS) is adequate (for more details see, \emph{e.g.}, Ref.\tcite{Alekhin:2009ni} and references therein). In the FFNS only light-flavored quarks and gluons are accounted for by proton collinear parton densities (PDFs). Moreover, heavy quarks are only produced in the final state and their masses cannot be neglected, otherwise power corrections proportional to $m_b^2/\vec{p}_T^{\; 2}$ would arise and spoil the convergence of perturbative series.
Conversely, at high $|\vec{p}_T|$ values, namely when $|\vec{p}_T| \gg m_b$, terms proportional to $\ln (|\vec{p}_T|/m_b)$ become larger and larger and must be resummed to all orders\tcite{Mele:1990cw,Cacciari:1993mq}. In the latter case, the so-called zero-mass variable-flavor number scheme (ZM-VFNS, or simply VFNS) is used to match NLO predictions with resummed calculations\tcite{Buza:1996wv,Bierenbaum:2009mv,Cacciari:1993mq,Binnewies:1997xq}. Here, all flavors are present in the initial state and are taken massless. A matching among ZM-VFNS and FFNS, aimed at blending the advantages of the two schemes, exists and it is known as general-mass variable-flavor number scheme (GM-VFNS). 
It combines massive (low scale) and massless (high scale) calculations, and the heavy-quark masses are used as parameters by which FFNS turns into VFNS.
Different implementations of this scheme have been proposed so far\tcite{Kramer:2000hn,Forte:2010ta,Blumlein:2018jfm,Aivazis:1993pi,Thorne:1997ga}, and for a detailed discussion we refer the reader to Ref.\tcite{Gao:2017yyd} (see also Refs.\tcite{Buza:1996wv,Bierenbaum:2009mv}).

In Refs.\tcite{Kniehl:2010iz,Saleev:2012np} inclusive bottom-jet emissions in central-rapidity regions were investigated under the hypothesis of $t$-channel exchanges of gluon and quark Reggeons at high energies. These studies were subsequently extended to bottomed bound states\tcite{Karpishkov:2014epa,Karpishkov:2016eaj,Karpishkov:2017kph}.
In Ref.\tcite{Maciula:2010yw} the kinematic correlations of lepton pairs from semileptonic decays of charmed and bottomed mesons were discussed.
In Ref.\tcite{Maciula:2018mig} the weight of \emph{double-parton scattering} effects was assessed in the hadroproduction of a $D^0 B^+$ system and of two $B^+$ mesons at the LHC.

The $B$-meson VFNS collinear fragmentation function (FF) was first extracted at NLO in Ref.\tcite{Binnewies:1998vm} from a fit to $e^+e^-$~data elaborated by the CERN LEP1 Collaboration. Then, the parametrization obtained in Ref.\tcite{Kniehl:2008zza} via a fit to CERN-LEP1 and SLAC-SLC data was used to calculate the NLO cross section for the inclusive production of $B$~mesons in $pp$ collisions and in the GM-VFNS, namely by taking into account finite-mass effects of the bottom quark\tcite{Kniehl:2011bk}.

In Ref.\tcite{Kramer:2018vde} the hadroproduction of bottom-flavored hadrons ($B$ mesons and $\Lambda_b$ baryons, comprehensively indicated as $b$-hadrons) was investigated at LHC energies and compared with CMS and LHCb data. This study was performed under the assumption that a unique FF can be adopted to describe the fragmentation of partons to all $b$-hadrons species. Thus, the FF set for a given species could be obtained from the global one by simply multiplying the latter by a branching fragmentation fraction, which does not depend on energy.
Analyses done by the Heavy Flavor Averaging Group (HFAG)\tcite{Amhis:2016xyh} have shown how the universality assumption on the branching fraction is violated by LEP and Tevatron data for $\Lambda_b$ emissions, while its safety is corroborated for $B$-meson detections. A recent study on transverse-momentum distributions for the inclusive $\Lambda_b$ production at CMS and LHCb\tcite{Kramer:2018rgb} has pointed out that the branching-fraction picture needs to be improved in the large $p_T$-regime, and future data with reduced experimental uncertainties are expected to better clarify the situation.

The NLO fragmentation of $c$~and~$\bar{b}$~quarks to $B_c^{(*)}$~mesons was studied in Ref.\tcite{Zheng:2019gnb}, while the first determination of a next-to-NLO $b$-hadron FF via a fit to $e^+e^-$~annihilation data from CERN LEP1 and SLAC SLC was presented in Ref.\tcite{Salajegheh:2019ach}.
Energy and angular distributions for $b$-hadrons production from semi-leptonic decays of top quarks were analyzed in Refs.\tcite{Kniehl:2012mn,Kniehl:2021qep}.

Apart from direct-production channels, $b$-quark emissions are employed to identify top particles and to study their properties. Thus, the $b$-quark fragmentation mechanisms is expected to have a relevant phenomenological impact on top physics.
The same formalism can be applied in electroweak precision studies to describe photon radiation from massive charged fermions, such as a Higgs-boson detection via the $b \bar{b}$ decays\tcite{Buonocore:2017lry}. The role of the $b$-quark in the associated production of a lepton pair was discussed in Refs.\tcite{Maltoni:2012pa,Bagnaschi:2018dnh,Lim:2016wjo}. The treatment of electroweak radiation from heavy fermions in the context of $W$-boson production with Monte Carlo generators was extensively investigated in Ref.\tcite{Barze:2012tt}.

The picture described above is still incomplete if we approach particular kinematic regions where the perturbative series is poorly convergent. A prominent example is represented by the Sudakov region, where the ratio $x_S$ between the transverse momentum of the detected particle and center-of-mass energy approaches one. Here, soft-gluon radiation produces contributions proportional to $\alpha_s^n \ln^m (1-x_S) / (1-x_S)$, with $m \leq 2n-1$, which must be resummed\tcite{Cacciari:2001cw,Mele:1990cw}. This is equivalent to saying that the ``true"
expansion parameter is $\alpha_s \ln^2 (1-x_S)$ instead of $\alpha_s$. 

A similar issue arises when one approaches the so-called \textit{semi-hard} region of QCD (see Section\tref{ssec:semi-hard} for further details), namely where the scale hierarchy $s \gg \{Q^2\} \gg \LQCD^2$ ($s$~is the squared center-of-mass energy, $\{Q^2\}$ one or a set of squared hard scales given by the process kinematics, and $\LQCD$ the QCD mass scale) stringently holds\footnote{\revdone{We adopted here the standard definition of ``semi-hard'' processes and stress that the prefix ``semi-'' does
not mean attenuation of the hardness, but rather that the hard scale(s) is(are) not as large as $s$, as in a
``hard process''. Indeed, the impact factors for a semi-hard process, to be defined later, can be calculated
perturbatively due to the hardness of the process in the fragmentation regions of the colliding particles.}}. The possibility of entering this two-scale regime via the heavy-flavor production was highlighted many years ago, when the so-called \textit{high-energy factorization} (HEF) was proposed\tcite{Catani:1990xk,Catani:1990eg,Catani:1994sq}. Here, $m_b$ plays the role of hard scale.

In this paper we investigate the inclusive semi-hard emission at the LHC of a $b$-hadron accompanied by another $b$-hadron or by a light jet, as a testfield for the manifestation of imprints of the QCD high-energy dynamics. We build predictions for distributions differential in rapidity, azimuthal angles and observed transverse momenta, calculated at the hand of a \emph{hybrid} factorization that combines the Balitsky--Fadin--Kuraev--Lipatov (BFKL) resummation \tcite{Fadin:1975cb,Kuraev:1976ge,Kuraev:1977fs,Balitsky:1978ic} of leading and next-to-leading energy logarithms with collinear PDFs and FFs.
We hunt for stabilizing effects of the high-energy series under higher-order corrections and energy-scale variation, that, if confirmed, would pave the way toward prospective studies where the use of our hybrid factorization could serve as an important tool to improve precision calculations of observables sensitive to bottom-flavored bound-state emissions.

\section{Inclusive $b$-hadron production}
\label{sec:theory}

\begin{figure*}[tb]
\centering
\includegraphics[width=0.45\textwidth]{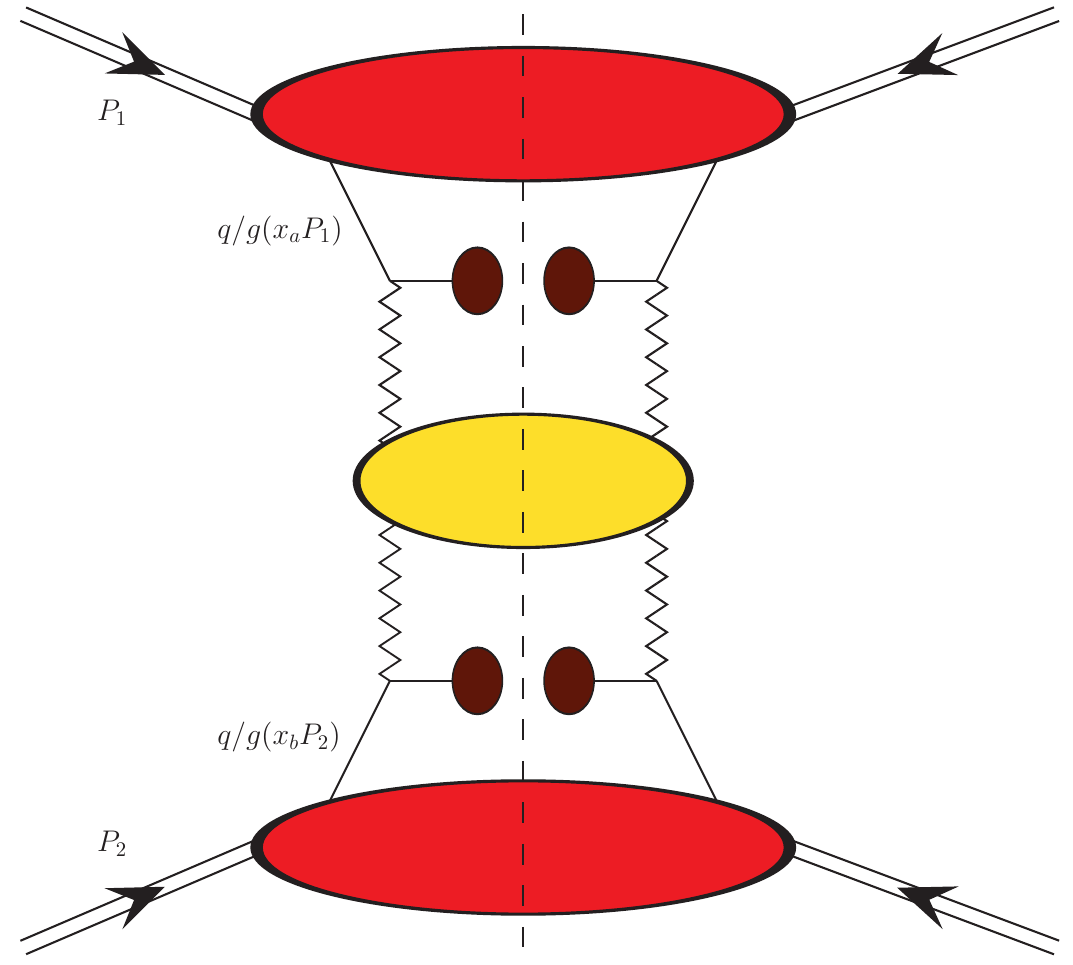}
\hspace{0.25cm}
\includegraphics[width=0.45\textwidth]{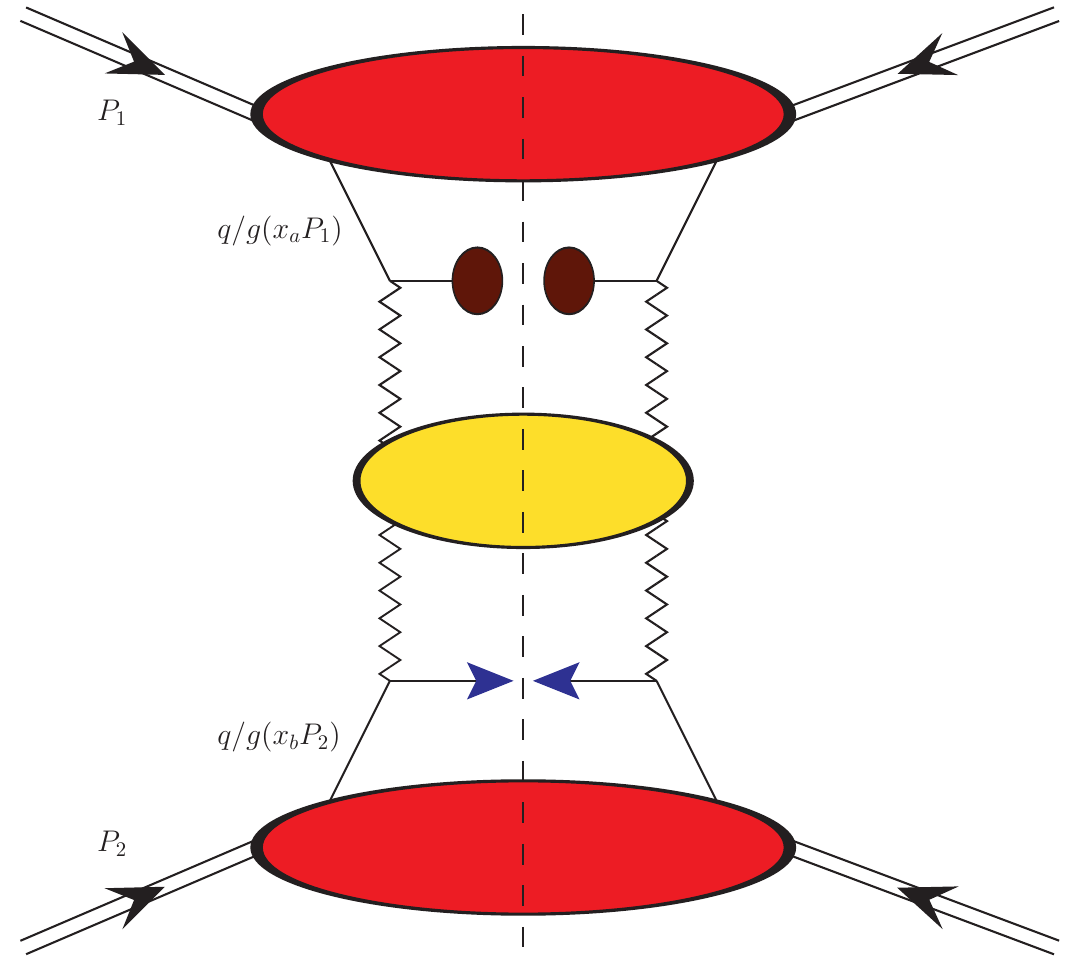}
\\ \vspace{0.25cm}
\hspace{-0.50cm}
a) Double $H_b$ \hspace{5.50cm}
b) $H_b$ $+$ jet
\caption{Hybrid high-energy/collinear factorization at work. Schematic representation of the two inclusive processes under investigation. Red blobs denote proton collinear PDFs, claret ovals depict $b$-hadron collinear FFs, and blue arrows stand for the light-jet selection algorithm. The BFKL ladder, represented by the yellow blob, is connected to impact factors via Reggeon (zigzag) lines. Diagrams were realized via the {\tt JaxoDraw 2.0} interface~\cite{Binosi:2008ig}.}
\label{fig:process}
\end{figure*}

In this Section we give theoretical key ingredients to build our observables. After a brief overview on recent progresses in the phenomenology of the semi-hard sector~(Section\tref{ssec:semi-hard}), we provide with analytic expressions of azimuthal-angle coefficients for our processes (see Fig.\tref{fig:process}), calculated in the hybrid high-energy and collinear factorization framework~(Section\tref{ssec:cross_section}). Then we present our choice for perturbative and non-perturbative ingredients~(Section\tref{ssec:ingredients}), as the running coupling, collinear PDFs and FFs, and the jet-algorithm selection. Finally, key features of the BLM scale optimization procedure are briefly shown~(Section\tref{ssec:BLM}).

\subsection{Semi-hard phenomenology at a glance}
\label{ssec:semi-hard}

As it is well known, the description of hadronic reactions at colliders has represented, and still represents, a great challenge for physicists. The possibility to decouple the long-distance dynamics from the short-distance one, and thus non-perturbartive ingredients from perturbative calculations via the well-known collinear factorization, is certainly one of the greatest achievements of modern particle physics. There exist, however, kinematic regimes which lie outside the domain of the standard collinear approach. This calls for an extension of the theoretical description that embodies the effect of one or more resummation mechanisms. 

In this work, our interest falls into the so-called \textit{semi-hard} sector,  where, as mentioned in Section\tref{sec:intro}, the scale hierarchy $s \gg Q^2 \gg \LQCD^2$ strictly holds. While the second inequality simply justifies the use of perturbation theory, the first tells us that we are in the so-called \emph{Regge limit} of QCD, where large logarithms of the ratio $s/Q^2$ enter the perturbative series with a power increasing together with the order. When $\alpha_s (Q^2) \ln (s/Q^2) \sim 1 $, a pure fixed-order perturbative calculation cannot provide with reliable predictions and a resummation to all orders, that catches the effect of these large logarithms, is needed. The most powerful framework for this resummation is the BFKL approach. This method prescribes how to resum all terms proportional to $(\alpha_s \ln s )^n$, the so called leading
logarithmic approximation (LLA), and all terms proportional to $\alpha_s(\alpha_s \ln s)^n$, the so called next-to-leading logarithmic approximation (NLA). In the BFKL approach, a generic scattering amplitude can be expressed as the convolution of a process-independent Green's function with two impact factors, related to the transition from each colliding particle to the respective final-state object. The BFKL Green’s function satisfies an integral equation, whose kernel is known at the NLO for any fixed (not growing with energy) momentum transfer, $t$, and for any possible two-gluon color configuration in the $t$-channel\tcite{Fadin:1998py,Ciafaloni:1998gs,Fadin:1998jv,Fadin:2000hu,Fadin:2004zq,Fadin:2005zj}. 

Despite the NLO accuracy achieved in the calculation of the kernel, the predictive power of the BFKL approach in its full NLA realization is limited by the number of impact factors known at NLO order: 1) colliding-parton (quarks and gluons) impact factors\tcite{Fadin:1999de,Fadin:1999df},
which represents the basis for constructing the 2) forward-jet impact factor\tcite{Bartels:2001ge,Bartels:2002yj,Caporale:2011cc,Caporale:2012ih,Ivanov:2012ms,Colferai:2015zfa} and 3) forward light hadron one\tcite{Ivanov:2012iv}, 4) the impact factor describing the $\gamma^*$ to light-vector-meson (LVM) leading twist transition\tcite{Ivanov:2004pp}, 5) the one detailing the $\gamma^* \rightarrow \gamma^*$ subprocess\tcite{Bartels:2000gt,Bartels:2001mv,Bartels:2002uz,Bartels:2004bi,Fadin:2001ap,Balitsky:2012bs}, and 6) the one for the production of a forward Higgs boson in the infinite top-mass limit\tcite{Hentschinski:2020tbi,Nefedov:2019mrg}. On the other hand, if we limit ourselves to the LLA accuracy, other impact factors calculated at leading-order (LO) can be considered: 7) forward Drell--Yan pair, 8) forward heavy-quark pair both in the hadroproduction and photoroduction channel\tcite{Celiberto:2017nyx,Bolognino:2019yls}, forward $J/\Psi$\tcite{Boussarie:2017oae}, 10) Higgs in the central region of rapidity\tcite{Lipatov:2005at,Pasechnik:2006du,Abdulov:2017tis,Lipatov:2014mja}. Some universal NLO corrections can be added to the LO impact factors, based on renormalization group analysis and on the invariance under variation of the energy scale $s_0$ entering the BFKL approach.

On one side, impact factors has been used to build up predictions for a considerable number of (inclusive) reactions featuring a forward-plus-backward two-particle final state. An incomplete list includes: the exclusive diffractive leptoproduction of two light vector mesons\tcite{Pire:2005ic,Segond:2007fj,Enberg:2005eq,Ivanov:2005gn,Ivanov:2006gt}, the inclusive hadroproduction of two jets featuring large transverse momenta and well separated in rapidity (Mueller--Navelet channel\tcite{Mueller:1986ey}), for which several phenomenological studies have appeared so far~(see, \emph{e.g.},~Refs.\tcite{Colferai:2010wu,Caporale:2012ih,Ducloue:2013hia,Ducloue:2013bva,Caporale:2013uva,Caporale:2014gpa,Colferai:2015zfa,Caporale:2015uva,Ducloue:2015jba,Celiberto:2015yba,Celiberto:2015mpa,Celiberto:2016ygs,Celiberto:2016vva,Caporale:2018qnm}), the inclusive detection of two light-charged rapidity-separated hadrons\tcite{Celiberto:2016hae,Celiberto:2016zgb,Celiberto:2017ptm}, three- and four-jet hadroproduction\tcite{Caporale:2015vya,Caporale:2015int,Caporale:2016soq,Caporale:2016vxt,Caporale:2016xku,Celiberto:2016vhn,Caporale:2016djm,Caporale:2016lnh,Caporale:2016zkc}, $J/\Psi$-plus-jet\tcite{Boussarie:2017oae}, hadron-plus-jet\tcite{Bolognino:2019cac}, Higgs-plus-jet\tcite{Celiberto:2020tmb,Celiberto:2021fjf}, heavy-light dijet system\tcite{Bolognino:2021mrc,Bolognino:2021hxx} and forward Drell–Yan dilepton production with a
possible backward-jet tag~\cite{Golec-Biernat:2018kem}.

On the other side, the study of single forward emissions offers us the possibility to probe the proton content via an \emph{unintegrated gluon distribution} (UGD), whose evolution in the struck-gluon longitudinal fraction $x$ is driven by BFKL. Probe candidates of the UGD are: the exclusive light vector-meson electroproduction~\cite{Anikin:2009bf,Anikin:2011sa,Besse:2013muy,Bolognino:2018rhb,Bolognino:2018mlw,Bolognino:2019bko,Bolognino:2019pba,Celiberto:2019slj,Bolognino:2021niq,Bolognino:2021gjm}, the exclusive quarkonium photoproduction\tcite{Bautista:2016xnp,Garcia:2019tne,Hentschinski:2020yfm}, and the inclusive tag of Drell--Yan pairs in forward directions\tcite{Motyka:2014lya,Brzeminski:2016lwh,Motyka:2016lta,Celiberto:2018muu}.
The information on the gluon content at small-$x$ embodied in the UGD turned out to be relevant in the improvement of the collinear description via a first determination of small-$x$ resummed PDFs\tcite{Ball:2017otu,Abdolmaleki:2018jln,Bonvini:2019wxf}, as well as in a model calculation of small-$x$ transverse-momentum-dependent gluon densities (TMDs)\tcite{Bacchetta:2020vty,Celiberto:2021zww,Bacchetta:2021oht}. Studies on the interplay between BFKL dynamics and TMD factorization were recently made in Refs.\tcite{Nefedov:2021vvy,Hentschinski:2021lsh}.

A major issue emerging in phenomenological applications of the BFKL approach to semi-hard observables is the fact that NLA corrections both to the BFKL Green's function and impact factors turn out to be of the same size and with opposite sign of pure LLA contributions. This makes the high-energy series unstable and this becomes strongly manifest when studies on renormalization/factorization scale variation are performed. More in particular, it was pointed out how BFKL-sensitive observables, such as azimuthal-angle correlations in the Mueller--Navelet reaction, cannot be studied at ``natural" scales\tcite{Ducloue:2013bva,Caporale:2014gpa,Celiberto:2020wpk}. A general procedure that allows us to ``optimize" scales in semi-hard final states was built up in Ref.\tcite{Caporale:2015uva}. It relies on the so-called Brodsky--Lepage--Mackenzie (BLM) method\tcite{Brodsky:1996sg,Brodsky:1997sd,Brodsky:1998kn,Brodsky:2002ka}, which prescribes that the optimal scale value is the one that cancels the non-conformal $\beta_0$-terms in the considered observable. Although the application of the BLM method led to a significant improvement of the agreement between predictions for azimuthal correlations of the two Mueller--Navelet jets and CMS data\tcite{Khachatryan:2016udy}, the scale values found, much higher than the natural ones, generally bring to a substantial reduction of cross sections (observed for the first time in inclusive light charged dihadron emissions\tcite{Celiberto:2016hae,Celiberto:2017ptm}). This issue clearly hampers any possibility of doing precision studies.

First, successful attempts at gaining stability of BFKL observables under higher-order corrections at natural scales were made via the analysis of semi-hard states featuring the detection of objects with large transverse masses, such as Higgs bosons\tcite{Celiberto:2020tmb} and heavy-flavored jets\tcite{Bolognino:2021mrc}. However, due to the lack of a NLO calculation for the corresponding impact factors (as mentioned before, the NLO Higgs impact factor was calculated quite recently in the large top-mass limit only), these reactions were studied with partial NLA accuracy. The first evidence of stabilizing effects in semi-hard processes studied at NLA came out in a recent study on inclusive $\Lambda_c$ emissions\tcite{Celiberto:2021dzy}. It was highlighted how the peculiar behavior of VFNS FFs depicting the baryon production at large transverse momenta\tcite{Kniehl:2020szu} acts as a fair stabilizer of high-energy predictions for observables sensitive to double $\Lambda_c$ final states, while a partial stabilization was found in the production of a $\Lambda_c$ particle plus a light-flavored jet. Further studies on other channels featuring the tag of heavier hadron species are thus needed to corroborate the statement that the heavy-flavor production is a suitable testing ground for the manifestation of the aforementioned stabilizing effects.

In this work we consider the inclusive semi-hard production of two $b$-hadrons of a $b$-hadron plus jet system,
\begin{equation}
\label{processes}
    p(P_1) + p(P_2) \rightarrow H_b(p_1, y_1) + X + H_b(p_2, y_2) \; ,
\end{equation}
\begin{equation*}
    p(P_1) + p(P_2) \rightarrow H_b(p_1, y_1) + X + {\rm{jet}} (p_2, y_2) \; ,
\end{equation*}
where $p(P_{1,2})$ stands for an initial proton with momenta $P_{1,2}$, $H_b(p_i, y_i)$ for a generic bottom-flavored hadron\footnote{In our analysis we are inclusive on the production of all species of $b$-hadrons whose lowest Fock state contains either a $b$ or $\bar b$ quark, but not both. Therefore, bottomed quarkonia are not considered. Furthermore, we ignore $B_c$ mesons since their production rate
is estimated to be at most 0.1\% of $b$-hadrons (see, \emph{e.g.}, Refs.\tcite{LHCb:2014iah,LHCb:2016qpe}). Our choice is in line with the $b$-hadron FF determination of Ref.\tcite{Kramer:2018vde}.} with momentum $p_i$ and rapidity $y_i$, and $X$ contains all the undetected products of the reaction. The semi-hard configuration is realized when the two detected objects possess large transverse masses, $m_{1,2 \perp} \gg \LQCD$, with $m_{1,2 \perp} = \sqrt{|\vec p_{1,2}|^2 + m_{1,2}^2}$\,, and $\vec p_{1,2}$ their transverse momenta. A large rapidity separation, $\DY = y_1 - y_2$, is required in order to consider our reactions as diffractive ones\footnote{\revdone{The use of ``diffractive'' for our inclusive process is justified
because the undetected hadronic activity is concentrated in the central region and summed over, thus leading, via the optical theorem, to differential cross sections (in the kinematic variable of the colliding particles' fragmentation regions) which take the same form as in truly diffractive processes, where there
is no activity at all in the central region.}}. We will let observed transverse momenta in ranges sufficiently large to ensure the validity of a VFNS description.

\subsection{High-energy resummed cross section}
\label{ssec:cross_section}

Final-state configurations that distinguish the two processes under consideration are schematically represented in Fig.\tref{fig:process}, where a \textit{b}-hadron~$(p_1,y_1)$ is emitted along with another \textit{b}-hadron or a jet~$(p_2,y_2)$, featuring a large rapidity separation, $\DY$, together with an undetected system of hadrons. For the sake of definiteness, we will consider the case where the rapidity of the first detected final-state object, $y_1$, is larger than the second one, $y_2$, so that $\DY$ is always positive, and the first object is forward while the second is backward.

The colliding protons' momenta $P_1$ and $P_2$ are taken as Sudakov basis vectors satisfying
$P_1^2= P_2^2=0$ and $2 (P_1\cdot P_2) = s$, so that the four-momenta of detected
objects can be decomposed as
\begin{equation}\label{sudakov}
p_{1,2} = x_{1,2} P_{1,2} + \frac{\vec p_{1,2}^{\,2}}{x_{1,2} s}P_{2,1} + p_{1,2\perp} \; ,
\end{equation}
\begin{equation*}
p_{1,2\perp}^2=-\vec p_{1,2}^{\,2}\;.
\end{equation*}

In the large-rapidity limit, the outgoing particle longitudinal momentum fractions, $x_{1,2}$, are connected to the respective rapidities through the relation
$y_{1,2}=\pm\frac{1}{2}\ln\frac{x_{1,2}^2 s}
{\vec p_{1,2}^2}$, 
so that one has $\drv y_{1,2}~=~\pm~\frac{\drv x_{1,2}}{x_{1,2}}$,
and $\DY=y_1-y_2=\ln\frac{x_1x_2 s}{|\vec p_1||\vec p_2|}$, where the
spatial part of the four-vector $p_{1\parallel}$ is taken positive.

Within the pure QCD collinear factorization, LO cross section for our two reactions in Eq.\eref{processes} is given as a convolution of the partonic hard-scattering factor with the parent-proton PDFs and the FFs describing outgoing objects. One has
\begin{widetext}
\begin{equation}
\label{sigma_collinear_HbHb}
\frac{\drv \sigma^{[pp \,\rightarrow\, H_b H_b]}_{\rm coll.}}{\drv x_1\drv x_2\drv ^2\vec p_1\drv ^2\vec p_2}
=\sum_{r,s=q,{\bar q},g}\int_0^1 \drv x_a \int_0^1 \drv x_b\ f_r\left(x_a\right) f_s\left(x_b\right)
\end{equation}
\[
\times\int_{x_1}^1\frac{\drv \beta_1}{\beta_1}\int_{x_2}^1\frac{\drv \beta_2}{\beta_2}D^{H_b}_{r}\left(\frac{x_1}{\beta_1}\right)D^{H_b}_{s}\left(\frac{x_2}{\beta_2}\right)
\frac{\drv {\hat\sigma}_{r,s}\left(\hat s\right)}
{\drv x_1\drv x_2\drv ^2\vec p_1\drv ^2\vec p_2}\;,
\]
\begin{equation}
\label{sigma_collinear_HbJ}
\begin{split}
\frac{\drv\sigma^{[pp \,\rightarrow\, H_b \, {\rm jet}]}_{\rm coll.}}{\drv x_1\drv x_2\drv ^2\vec p_1\drv ^2\vec p_2}
=\sum_{r,s=q,{\bar q},g}\int_0^1 \drv x_a \int_0^1 \drv x_b\ f_r\left(x_a\right) f_s\left(x_b\right)
\int_{x_1}^1\frac{\drv \beta_1}{\beta_1}D^{H_b}_{r}\left(\frac{x_1}{\beta_1}\right) 
\frac{\drv {\hat\sigma}_{r,s}\left(\hat s\right)}
{\drv x_1\drv x_2\drv ^2\vec p_1\drv ^2\vec p_2}\;.
\end{split}
\end{equation}
\end{widetext}
In Eqs.\eref{sigma_collinear_HbHb}~and\eref{sigma_collinear_HbJ} the $r, s$ indices specify the parton types (quarks $q = u, d, s, c, b$; antiquarks $\bar q = \bar u, \bar d, \bar s, \bar c, \bar b$; or gluon $g$), $f_{r,s}\left(x, \mu_F \right)$ and $D^{H_b}_{r,s}\left(x/\beta, \mu_F \right)$ denote the initial proton PDFs and the final detected $b$-hadron FFs, respectively; $x_{a,b}$ are the longitudinal fractions of the partons involved in the hard subprocess and $\beta_{1(,2)}$ the longitudinal fraction(s) of the parton(s) fragmenting into $b$-hadron(s); $\drv\hat\sigma_{r,s}\left(\hat s \right)$ is
the partonic cross section and $\hat s \equiv x_a x_b s$ is the squared center-of-mass energy of the parton-parton collision subprocess.
For the sake of simplicity, the explicit dependence of PDFs, FFs and of the partonic cross section on the factorization scale, $\mu_F$, has been dropped.

Contrariwise to the pure collinear treatment, we build the cross section in hybrid factorization, where the high-energy dynamics is genuinely provided by the BFKL approach, and collinear ingredients are then embodied.
We decompose the cross section as a Fourier sum of azimuthal-angle coefficients, ${\cal C}_n$, in the following way
\begin{equation}
\hspace{-0.20cm}
\frac{(2\pi)^2 \drv \sigma}
{\drv y_{1,2} \drv |\vec p_{1,2}| \drv \phi_{1,2}}
=
\left[{\cal C}_0+\sum_{n=1}^\infty 2\cos (n\varphi)\,
{\cal C}_n\right]\, ,
 \label{dsigma_Fourier}
\end{equation}
where $\varphi=\phi_1-\phi_2-\pi$, with $\phi_{1,2}$ the outgoing particle azimuthal angles. 
In the NLA accuracy and in the $\MSb$ renormalization scheme\tcite{PhysRevD.18.3998} the $\varphi$-summed cross section, ${\cal C}_0$, and the other coefficients, ${\cal C}_{n > 0}$, are given by (for details on the derivation see, \emph{e.g.},~Ref.\tcite{Caporale:2012ih} \revdone{and in Ref.\tcite{Caporale:2015uva}})
\begin{widetext}
\vspace{-0.50cm}
\begin{equation}\label{Cn_NLA_NS_MSb}
\begin{aligned}
{\cal C}_n & \equiv \int_0^{2\pi}\drv \phi_1 \int_0^{2\pi}\drv \phi_2\,
\cos (n\varphi) \,
\frac{\drv \sigma}{\drv y_1\drv y_2\drv |\vec p_1|\drv |\vec p_2|\drv \phi_1\drv \phi_2}\;
\\ & = \, 
\frac{e^{\DY}}{s}
\int_{-\infty}^{+\infty}\drv \nu \, \left(\frac{x_a x_b s}{s_0}
\right)^{\bar \alpha_s(\mu_R)\left\{\chi(n,\nu)+\bar\alpha_s(\mu_R)
	\left[\bar\chi(n,\nu)+\frac{\beta_0}{8 N_c}\chi(n,\nu)\left[-\chi(n,\nu)
	+\frac{10}{3}+2\ln\left(\frac{\mu_R^2}{\sqrt{\vec p_1^2\vec p_2^2}}\right)\right]\right]\right\}}
\\ & \times 
\, \as^2(\mu_R) \, c_1(n,\nu,|\vec p_1|, x_1) \, 
[c_2(n,\nu,|\vec p_2|,x_2)]^*\,
\\ & \times \, 
\left\{1
+\as(\mu_R)\left[\frac{c_1^{(1)}(n,\nu,|\vec p_1|,
	x_1\textcolor{red}{,s_0})}{c_1(n,\nu,|\vec p_1|, x_1)}
+\left[\frac{c_2^{(1)}(n,\nu,|\vec p_2|, x_2\textcolor{red}{,s_0})}{c_2(n,\nu,|\vec p_2|,
	x_2)}\right]^*\right] + \bar\alpha_s^2(\mu_R) \DY
\frac{\beta_0}{4 N_c}\chi(n,\nu)f(\nu)\right\}\;.
\end{aligned}
\end{equation}
\end{widetext}
Here $\bar \alpha_s(\mu_R) \equiv \alpha_s(\mu_R) N_c/\pi$, with $N_c$ the number of colors, $\beta_0$ is the first coefficient of the QCD $\beta$-function (see Eq.\eref{as_parameters}),
\begin{equation}
\hspace{-10pt}
\chi\left(n,\nu\right)=2\left\{\psi\left(1\right)-{\rm Re} \left[\psi\left(\frac{n+1}{2}+i\nu \right)\right] \right\}
\label{chi_BFKL}
\end{equation}
is the LO BFKL characteristic function,
\revdone{
while
$\bar\chi(n,\nu)$, calculated in Ref.~\cite{Kotikov:2000pm} (see also Ref.~\cite{Kotikov:2002ab}), is the NLO correction to the BFKL kernel
\begin{widetext}
\[
 \bar \chi(n,\nu)\,=\, - \frac{1}{4}\left\{\frac{\pi^2 - 4}{3}\chi(n,\nu) - 6\zeta(3) - \chi^{\prime\prime}(n,\nu)  + \,2\,\phi(n,\nu) + \,2\,\phi(n,-\nu)
\right.
\]
\begin{equation}
 \label{kernel_NLO}
 \left.
 + \; \frac{\pi^2\sinh(\pi\nu)}{2\,\nu\, \cosh^2(\pi\nu)}
 \left[
 \left(3+\left(1+\frac{n_f}{N_c^3}\right)\frac{11+12\nu^2}{16(1+\nu^2)}\right)
 \delta_{n0}
 -\left(1+\frac{n_f}{N_c^3}\right)\frac{1+4\nu^2}{32(1+\nu^2)}\delta_{n2}
\right]\right\} \, ,
\end{equation}
with
\[
 \phi(n,\nu)\,=\,-\int\limits_0^1 \drv x\,\frac{x^{-1/2+i\nu+n/2}}{1+x}\left\{\frac{1}{2}\left(\psi^\prime\left(\frac{n+1}{2}\right)-\zeta(2)\right)+\mbox{Li}_2(x)+\mbox{Li}_2(-x)\right.
\]
\[
\left.
 +\; \ln x\left[\psi(n+1)-\psi(1)+\ln(1+x)+\sum_{k=1}^\infty\frac{(-x)^k}{k+n}\right]+\sum_{k=1}^\infty\frac{x^k}{(k+n)^2}\left[1-(-1)^k\right]\right\}
\]
\[
 = \; \sum_{k=0}^\infty\frac{(-1)^{k+1}}{k+(n+1)/2+i\nu}\left\{\psi^\prime(k+n+1)-\psi^\prime(k+1)\right.
\]
\begin{equation}
\label{kernel_NLO_phi}
 \left.
 + \; (-1)^{k+1}\left[\beta_{\psi}(k+n+1)+\beta_{\psi}(k+1)\right]-\frac{\psi(k+n+1)-\psi(k+1)}{k+(n+1)/2+i\nu}\right\} \; ,
\end{equation}
\begin{equation}
\label{kernel_NLO_phi_beta_psi}
 \beta_{\psi}(z)=\frac{1}{4}\left[\psi^\prime\left(\frac{z+1}{2}\right)
 -\psi^\prime\left(\frac{z}{2}\right)\right] \;
\end{equation}
\end{widetext}
and
\begin{equation}
\label{dilog}
\mbox{Li}_2(z) = - \int\limits_0^x \drv x \,\frac{\ln(1-x)}{x} \; .
\end{equation}
Then,
}$c_{1,2}(n,\nu)$ are the LO forward/backward objects impact factors in the
$(n,\nu)$-repre\-sen\-ta\-tion, whose compact expression for both the $H_{b}$ particle and the jet reads 
%
\begin{widetext}
\begin{equation}
c_{i}(n,\nu,|\vec p^{\,}|,x) = 2 \sqrt{\frac{C_F}{N_c}}
(\vec p^{\,2})^{i\nu-1/2}\,\int_{x}^1\drv \beta
\left( \frac{\beta}{x}\right)^{2 i\nu-1}
\label{LOIF}
\left[\frac{C_A}{C_F}f_g(\beta) \mathcal{S}_g^{(i)}(x,\beta)
+\sum_{r=q,\bar q}f_r(\beta) \mathcal{S}_r^{(i)}(x,\beta)\right] \;,
\end{equation}
\end{widetext}
where
\begin{equation}
\label{s_func}
\mathcal{S}_{g,r}^{(i)}(x,\beta) = \left\{
\begin{aligned}
&  \frac{1}{\beta} D_{g,r}^{H_b}(x/\beta) \; , 
\qquad & i = b\text{-hadron} \; , \\ 
& \delta(\beta-x) \; ,
\qquad & i = \text{jet}  \; , 
\end{aligned}
\right.
\end{equation}
and the $f(\nu)$ function is defined as
\begin{equation}
\label{fnu}
i\frac{\drv}{\drv\nu}\ln \frac{c_1}{[c_2]^*} =2\left[f(\nu)
-\ln\left(\sqrt{|\vec p_1| |\vec p_2|}\right)\right] \;.
\end{equation}
The remaining objects are the NLO corrections to impact factor in the Mellin representation
(also known as ($\nu, n$)-representation), $c_i^{(1)}(n,\nu,|\vec p_i|, x_i\textcolor{red}{,s_0})$.
As for the $H_b$ NLO impact factor, we rely on a light-hadron calculation, done in Ref.\tcite{Ivanov:2012iv}. This choice is consistent with our VFNS treatment, provided that energy scales at work are much larger than the bottom mass (see Section\tref{sec:results}).
Our selection for the light-jet NLO impact factor is discussed in Section\tref{ssec:ingredients}.

The way our hybrid factorization is realized fairly emerges from Eqs.~(\ref{Cn_NLA_NS_MSb})~and~(\ref{LOIF}). Here, azimuthal coefficients are high-energy factorized as convolutions of the gluon Green's function and the impact factors. The latter ones embody collinear ingredients, namely PDFs and FFs.
It is possible to obtain the LLA limit of our coefficients in Eq.~(\ref{Cn_NLA_NS_MSb}) by keeping just the LO part of the exponentiated kernel and by setting at zero the NLO impact factor corrections.

We employ NLA expressions given in this Section at the \emph{natural} energy scales given by the considered final state, \emph{i.e.} we set $\mu_R = \mu_F = \mu_N \equiv \sqrt{m_{1 \perp} m_{2 \perp}}$, where $m_{i \perp}$ is the transverse mass of the $i$-th emitted particle. Thus, one always have $m_{1 \perp} = \sqrt{|\vec p_1|^2 + m_{H_b}^2}$. Then, $m_{2 \perp} = \sqrt{|\vec p_2|^2 + m_{H_b}^2}$ in the double $H_b$ channel, whereas $m_{2 \perp}$ coincides with the jet transverse momentum in the $H_b$~$+$~jet one. We set $m_{H_b} = m_{\Lambda_b} = 5.62$ GeV, that corresponds to the mass of the heaviest $b$-hadron considered in our study. The $s_0$ energy scale is set equal to $\mu_N$.
  
\subsection{Perturbative and non-perturbative ingredients}
\label{ssec:ingredients}

In our calculations a two-loop running-coupling setup with $\alpha_s\left(M_Z\right)=0.11707$ and $n_f=5$ is adopted. Its $\MSb$-scheme expression is
\begin{equation}
\label{as_MSb}
 \as(\mu_R) \equiv \as^{\MSb}(\mu_R) = \frac{\pi}{\beta_0 L_R} \left( 4 - \frac{\beta_1}{\beta_0^2} \frac{\ln L_R}{L_R} \right) \;,
\end{equation}
with
\begin{equation}
\label{as_parameters}
 L_R(\mu_R) = 2 \ln \frac{\mu_R}{\LQCD} \;,
\end{equation}
\[
 \beta_0 = 11 - \frac{2}{3} n_f \;, \quad
 \beta_1 = 102 - \frac{38}{3} n_f \;.
\]
We introduce here also the MOM renormalization scheme\tcite{Barbieri:1979be,PhysRevD.20.1420,PhysRevLett.42.1435}, because this is the scheme in which the BLM procedure is developed (see section \ref{ssec:BLM}).
The MOM-scheme expression of the strong coupling, $\alpha_s^{\rm{MOM}}$, is obtained by inverting the relation
\begin{equation}
\label{as_MOM}
 \as^{\MSb} = \as^{\rm MOM} \left( 1 +  \frac{\tau^{\beta} + \tau^{\rm conf}}{\pi} \as^{\rm MOM} \right) \;,
\end{equation}
with
\begin{equation}
\label{T_bc}
\tau^\beta = - \left( \frac{1}{2} + \frac{I}{3} \right) \beta_0
\end{equation}
and
\begin{equation}
\label{T_conf}
\begin{aligned}
\tau^{\rm conf} & = \frac{C_A}{8}\left[ \frac{17}{2}I +\frac{3}{2}\left(I-1\right)\xi \right. \\
 & \left. + \, \left( 1-\frac{1}{3}I\right)\xi^2-\frac{1}{6}\xi^3 \right] \; ,
\end{aligned}
\end{equation}
where $C_A \equiv N_c$ is the color factor associated to a gluon emission from a gluon, then we have $I=-2\int_0^1\drv y \frac{\ln y}{y^2-y+1} \simeq 2.3439$, with the gauge parameter $\xi$  fixed at zero in the following.

It is well known that potential sources of uncertainty are expected to arise from the particular choice of the PDF parameterization. We performed preliminary tests on our observables by using the three most popular NLO PDF sets ({\tt MMHT14}\tcite{Harland-Lang:2014zoa}, {\tt CT14}\tcite{Dulat:2015mca} and {\tt NNPDF3.0}\tcite{Ball:2014uwa}), proving that PDF selection does not lead to a significant discrepancy in the kinematic regions of our interest. Furthermore, recent studies done via the so-called replica method\tcite{Forte:2002fg} have confirmed that BFKL-related observables, such as azimuthal correlations, exhibit a weak sensitivity to PDF replicas (see Section~3.3 of Ref.\tcite{Celiberto:2020wpk}). Therefore, in our analysis we employed the central value of an individual NLO PDF set, namely the {\tt MMHT14} one.

We depicted the parton fragmentation to $b$-hadrons by the hand of the {\tt KKSS07} NLO~FFs, that, as mentioned in the Introduction~(Section\tref{sec:intro}), were originally extracted from data of inclusive $B$-meson emissions in $e^+e^-$~annihilation\tcite{Kniehl:2008zza}.
In this parametrization the $b$ flavor has its starting scale at $\mu_0 = 4.5 \text{ GeV} \simeq m_b$ and is portrayed by a simple, three-parameter power-like \emph{Ansatz}\tcite{Kartvelishvili:1985ac}
\begin{equation}
\label{KKSS19_FF_power}
 D^{H_b}(x, \mu_0) = {\cal N} x^a (1-x)^b \;,
\end{equation}
whereas gluon and lighter quark (including $c$) FFs are generated
through DGLAP evolution and vanish at $\mu_F = \mu_0$.
Following Ref.\tcite{Kramer:2018vde}, we obtained the $b$-hadron FFs from the $B$-meson ones by simply removing the branching fraction for the $b \to B^\pm$ transition, which was assumed as $f_u = f_d = 0.397$ (see also Ref.\tcite{Kniehl:2008zza}). We stress that this choice is justified by the assumption that a unique FF can be adopted to describe the fragmentation of partons to all $b$-hadrons species, except for $\Lambda_b$ baryons.
We compared our predictions for $b$-hadrons' cross section with corresponding results for $\Lambda_c$~baryons and $\Lambda$~hyperons (see Appendix\tild\hyperlink{app:A}{A}) by respectively using {\tt KKSS19}\tcite{Kniehl:2020szu} and {\tt AKK08}\tcite{Albino:2008fy} NLO~FFs, which are close in the extraction technology to the {\tt KKSS07} set.

When the $H_b$~$+$~jet production channel is considered at NLA, a choice for the jet reconstruction algorithm, that enters the definition of the NLO jet impact factor, has to be made. The most popular classes of jet-selection functions are the $\kappa_\perp$~\emph{sequential-clustering}\tcite{Catani:1993hr} and the \emph{cone-type} algorithms\tcite{Ellis:1990ek}. 
A simpler version, \emph{infrared-safe} up to NLO perturbative accuracy and suited to numerical computations was derived in Ref.~\cite{Ivanov:2012ms} in the so-called ``small-cone'' approximation (SCA)~\cite{Furman:1981kf,Aversa:1988vb}, namely for a small-jet cone aperture in the rapidity-azimuthal angle plane.
Analytic expressions for the SCA jet vertex where then calculated in Ref.~\cite{Colferai:2015zfa} for both the $\kappa_\perp$ and the cone jet algorithms. Preliminary tests have shown that the adoption of these two versions allows for a reduction of the discrepancy between NLA corrections and pure LLA prediction for the $H_b$~$+$~jet cross section, with respect to the use of the original SCA algorithm of Ref.~\cite{Ivanov:2012ms}. Conversely, the cone-type SCA algorithm allows for a slighter stabilization of the NLA azimuthal-angle correlations under scale variation with respect to the original SCA one, while the $\kappa_\perp$ one leads to a stabilization worsening. Therefore, in our analysis on $H_b$~$+$~jet observables we used the cone-type SCA jet vertex with the jet-cone radius fixed at $R_J = 0.5$, as commonly done in recent experimental analyses at the LHC\tcite{Khachatryan:2016udy}. We postpone to a future work the dedicated study of all the systematic effects coming from the choice of the jet selection function in and beyond the SCA approximation.

\phantom{.} \vspace{0.0cm}

\subsection{BLM prescription on energy scales}
\label{ssec:BLM}

To test the stability of our observables under higher-order corrections and scale variation, we compare predictions at natural scales (see Section\tref{ssec:cross_section}) with the ones obtained by applying the BLM optimization method.
It essentially consists in finding the \emph{optimal} $\mu_R$ value, indicated as $\mu_R^{\rm BLM}$, as the value that removes all the non-conformal, $\beta_0$-dependent terms of the observable under consideration.
\revdone{In Ref.\tcite{Caporale:2015uva} a dedicated procedure was set up to remove all the non-conformal terms that appear in a given azimuthal coefficient $C_n$, namely $\beta_0$-dependent factors which appear \emph{both} in the NLA BFKL Green's function and in the NLO process-dependent impact factors}. This leads to a non-universality of the BLM scale and to its dependence on the energy of the process (and therefore on $\DY$).

Working in the MOM renormalization scheme,\revdone{in which the BLM procedure is natively implemented,} the optimal scale for a given azimuthal coefficient, $C_n$, is the value of $\mu_R$ that satisfies the condition
\begin{widetext}
\begin{equation}
\label{Cn_beta0_int}
  C_n^{(\beta_0)}(s, \DY) = 
  \int \drv \Phi(y_{1,2}, |\vec p_{1,2}|, \DY) \,
  \, {\cal C}_n^{(\beta_0)}  = 0 \, ,
\end{equation}
where $\drv \Phi(y_{1,2}, |\vec p_{1,2}|, \DY)$ stands for the final-state differential phase space (see Section\tref{sec:results}),
\[
 {\cal C}^{(\beta_0)}_n
 \propto \!\!
 \int_{-\infty}^{\infty} \drv\nu\ \left(\frac{x_a x_b s}{s_0}
\right)^{\bar \alpha^{\rm MOM}_s(\mu^{\rm BLM}_R)\chi(n,\nu)}
 c_1(n,\nu,|\vec p_1|,x_1)\,[c_2(n,\nu,|\vec p_2|,x_2)]^*
\]
\begin{equation}
\label{Cn_beta0}
 \times \, \left[{\omega}(\nu) + \bar \alpha^{\rm MOM}_s(\mu^{\rm BLM}_R) \DY \: \frac{\chi(n,\nu)}{2} \left(- \frac{\chi(n,\nu)}{2} + {\omega}(\nu) \right) \right] \, ,
\end{equation}
and
\begin{equation}
\label{upsilon_nu}
{\omega}(\nu) = f(\nu) - \frac{1}{3} (4 I + 1) + 2 \ln \left( \frac{\mu^{\rm BLM}_R}{\sqrt{|\vec p_1| |\vec p_2|}} \right) \, .
\end{equation}
\end{widetext}
\revdone{We remark that Eq.\eref{Cn_beta0} contains all the non-conformal terms present in Eq.\eref{Cn_NLA_NS_MSb} up to NLA accuracy.}
We define the scale ratio $ C_{\mu}^{\rm BLM} \equiv \mu_R^{\rm BLM}/\mu_N$, and look for the values of $C_{\mu}^{\rm BLM}$ which solve Eq.~(\ref{Cn_beta0_int}).
Then, the BLM scale value is plugged into formul{\ae} of the integrated coefficients, thus obtaining the following NLA BFKL expression in the MOM renormalization scheme
\begin{widetext}
\begin{equation}
\label{Cn_NLA_BLM_MOM_int}
\begin{aligned}
C_n^{\rm BLM\text{-}MOM} & = 
 \int \drv \Phi(y_{1,2}, |\vec p_{1,2}|, \DY) \; 
 \frac{e^{\DY}}{s} 
 \int_{-\infty}^{+\infty} \drv \nu \,
 \left(\alpha^{\rm MOM}_s (\mu^{\rm BLM}_R)\right)^2 \\
 & \times \,
 \left(\frac{x_a x_b s}{s_0}
\right)^{\bar \alpha^{\rm MOM}_s(\mu^{\rm BLM}_R)\left[\chi(n,\nu)
 +\bar \alpha^{\rm MOM}_s(\mu^{\rm BLM}_R)\left(\bar \chi(n,\nu) +\frac{\tau^{\rm conf}}
 {3}\chi(n,\nu)\right)\right]} \\
 & \times \,
 c_1(n,\nu,|\vec p_1|,x_1)[c_2(n,\nu,|\vec p_2|,x_2)]^* \\
 & \times \,
 \left\{1 + \alpha^{\rm MOM}_s(\mu^{\rm BLM}_R)\left[\frac{\bar c_1(n,\nu,|\vec p_1|,x_1\textcolor{red}{,s_0})}{c_1(n,\nu,|\vec p_1|,x_1)}
 +\left[\frac{\bar c_2(n,\nu,|\vec p_2|, x_2\textcolor{red}{,s_0})}{c_2(n,\nu,|\vec p_2|,x_2)}\right]^*
 +\frac{2}{\pi}\tau^{\rm conf} \right] \right\} \, ,
\end{aligned}
\end{equation}
\end{widetext}
where $\bar c_{{1,2}}(n,\nu,|\vec p_{1,2}|,x_{1,2}\textcolor{red}{,s_0})$ are the NLO impact-factor corrections after subtracting the non-conformal terms, which can be universally expressed through the LO impact factors, $c_{1,2}$. One has
\begin{equation}
\label{IF_NLO_sub}
\bar c_{1,2} =  c_{1,2}^{(1)} - \frac{\beta_0}{4 N_c} \left[ \pm i \frac{\drv}{\drv \nu} c_{1,2} + \left( \ln \mu_R^2 + \frac{5}{3} \right) c_{1,2} \right] \, .
\end{equation}
\revdone{In order to compare predictions at natural scales (Eq.\eref{Cn_NLA_NS_MSb}) with BLM-optimized results in the same renormalization scheme, we need to get the corresponding expression of Eq.\eref{Cn_NLA_BLM_MOM_int} in the $\MSb$ scheme.
This can be achieved by performing the two following replacements in Eq.\eref{Cn_NLA_BLM_MOM_int}
\begin{equation}
 \label{MOM_2_MSb}
 \alpha_s^{\rm MOM}(\mu^{\rm BLM}_R) \,\to\, \alpha_s^{\MSb}(\mu^{\rm BLM}_R) \;,
\end{equation}
\[
 \tau^{\rm conf} \,\to\, - \tau^\beta \;.
\]
In particular, we replace the analytic expression of the strong coupling in the MOM scheme, which is obtained by inverting Eq.\eref{as_MOM}, with the corresponding $\MSb$-one~(Eq.\eref{as_MSb}), while the value of $\mu_R$ is left unchanged.}

\section{Results and discussion}
\label{sec:results}

We present predictions for our observables that can be compared with forthcoming experimental analyses at 13~TeV LHC. Results for cross sections and azimuthal correlations, differential in the final-state rapidity distance, $\DY$, are discussed in Sections\tref{ssec:C0} and\tref{ssec:Rnm}, respectively. In Section\tref{ssec:pT} we introduce a new observable, namely the double differential $p_T$-distribution at fixed $\DY$, that can serve as a common basis for prospective studies on the interplay of different kinds of resummation mechanisms. Finally, a discussion on the stabilizing effects that our distributions gain when $b$-flavor FFs are considered is given in Appendix\tild\hyperlink{app:A}{A}.

The numerical elaboration of all the considered observables was done by making use of the {\tt JETHAD} modular work package\tcite{Celiberto:2020wpk}.
The sensitivity of our results on scale variation was assessed by letting $\mu_R$ and $\mu_F$ to be around their \emph{natural} values or their BLM \emph{optimal} ones, up to a factor ranging from 1/2 to two.
The $C_{\mu}$ parameter entering plots represents the ratio $C_\mu = \mu_{R,F}/\mu_N$. Error bands in our figures embody the combined effect of scale variation and phase-space multi-dimensional integration, the latter being steadily kept below 1\% by the {\tt JETHAD} integrators.
All calculations of our observables were done in the $\MSb$ scheme. BLM scales are calculated by solving the integral equation\eref{Cn_beta0_int} in the MOM scheme.

\subsection{$\DY$-distribution}
\label{ssec:C0}

The first observable under investigation is the cross section differential in the rapidity interval, also known as $\DY$-distribution or simply $C_0$. Its expression can be obtained by integrating the ${\cal C}_0$ azimuthal coefficient (see Eq.\eref{Cn_NLA_NS_MSb}) over transverse momenta and rapidities of the two final-state particles, and keeping $\DY$ fixed
\begin{widetext}
\begin{equation}
 \label{DY_distribution}
 C_0 =
 \int_{y_1^{\rm min}}^{y_1^{\rm max}} \drv y_1
 \int_{y_2^{\rm min}}^{y_2^{\rm max}} \drv y_2
 \int_{p_1^{\rm min}}^{p_1^{\rm max}} \drv |\vec p_1|
 \int_{p_2^{\rm min}}^{p_2^{\rm max}} \drv |\vec p_2|
 \, \,
 \delta (\DY - (y_1 - y_2))
 \, \,
 {\cal C}_0\left(|\vec p_1|, |\vec p_2|, y_1, y_2 \right)
 \, .
\end{equation}
\end{widetext}
The light-flavored jet is always tagged in its typical CMS ranges\tcite{Khachatryan:2016udy}, \emph{i.e.} $|y_J| < 4.7$ and 35~GeV~$< p_J <$~60~GeV.
A realistic proxy for the rapidity range of $b$-hadrons detected at the LHC could come from a recent study on $\Lambda_b$ baryons at CMS\tcite{Chatrchyan:2012xg}, $|y_{\Lambda_b}| < 2$. In our analysis we admit a tagging of $b$-hadrons on a slightly wider range, namely the one covered by the CMS barrel detector, $|y_H| < 2.4$. At variance with previous works, where lighter hadrons were studied in a transverse-momentum window from 10 to around 20 GeV (see, \emph{e.g.} Refs.\tcite{Celiberto:2017ptm,Celiberto:2020rxb,Celiberto:2021dzy}), here we allow the $b$-hadron transverse momentum to be in the range 20~GeV~$< |\vec p_H| <$~60~GeV, which is similar to the light-jet one. With this choice the validity of our VFNS treatment is preserved, since energy scales will be always much larger than the threshold for DGLAP evolution of the $b$-quark in {\tt KKSS07} FFs see Section\tref{ssec:ingredients} for more details).

\begin{figure*}[b]

   \includegraphics[scale=0.53,clip]{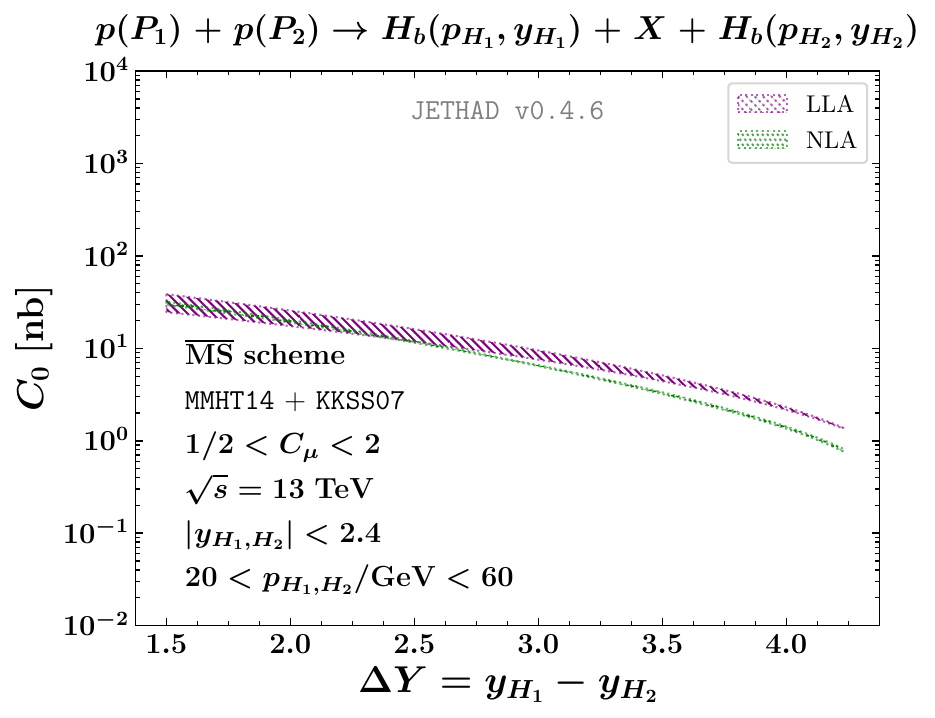}
   \includegraphics[scale=0.53,clip]{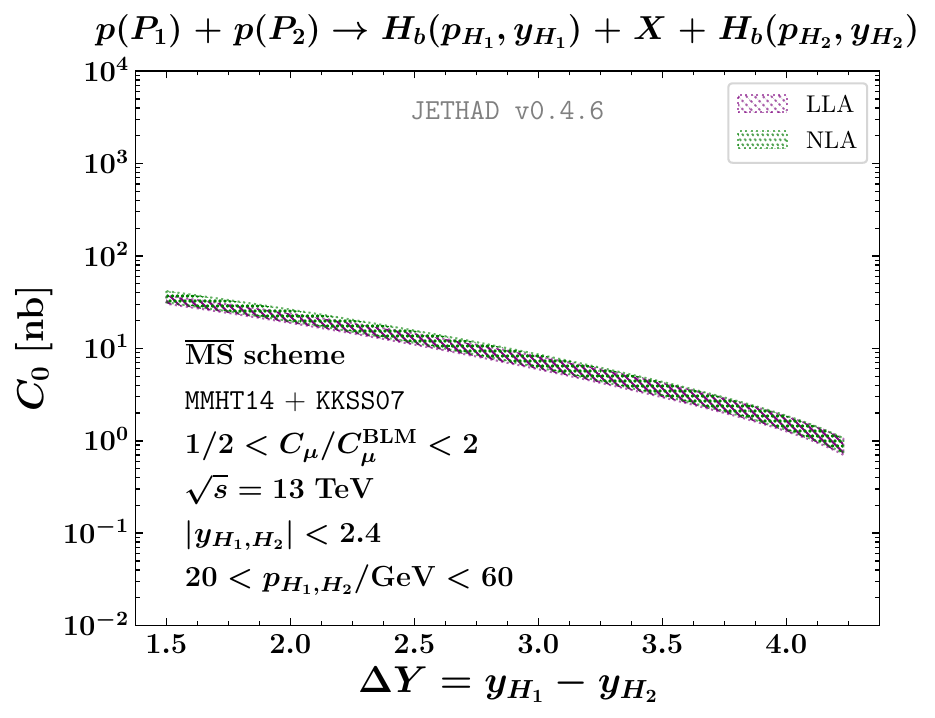}

   \includegraphics[scale=0.53,clip]{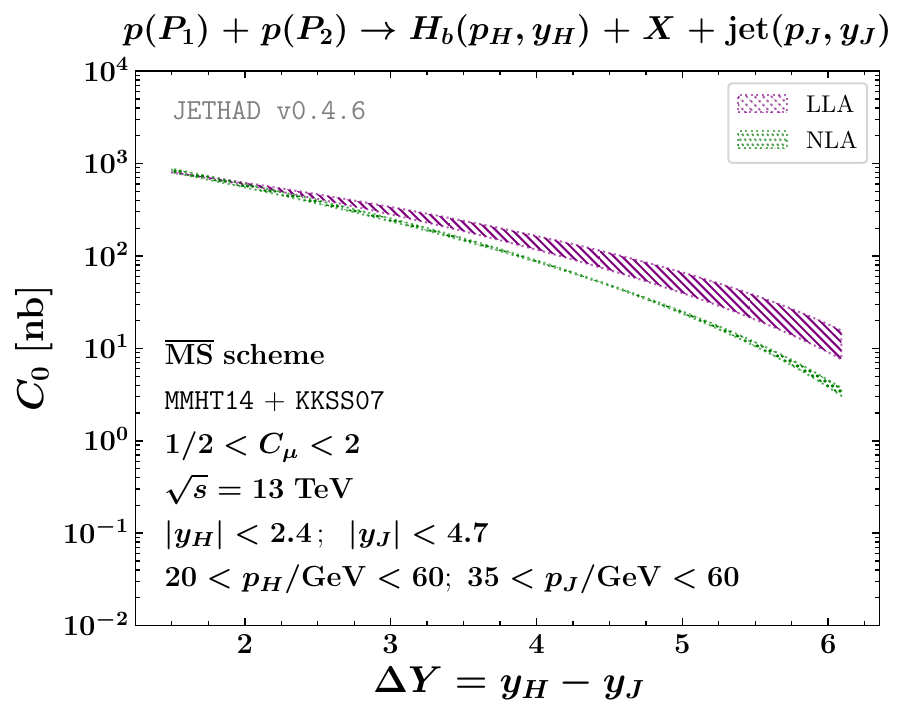}
   \hspace{0.10cm}
   \includegraphics[scale=0.53,clip]{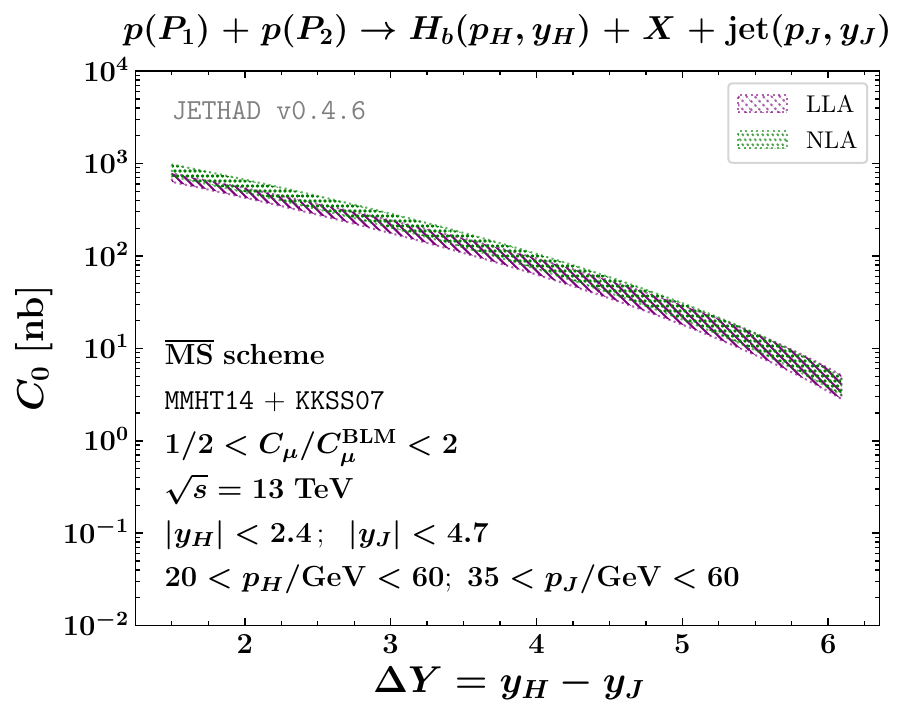}

\caption{$\DY$-shape of $C_0$ in the double $H_b$ (upper) and in the $H_b$~$+$~jet channel (lower), at natural (left) and BLM-optimized scales (right), and for $\sqrt{s} = 13$ TeV. Text boxes inside panels show transverse-momentum and rapidity ranges. Uncertainty bands embody the combined effect of scale variation and phase-space multi-dimensional integration.}
\label{fig:C0}
\end{figure*}

The $\DY$-shape of the $C_0$ distribution for the double $H_b$ production and for the $H_b$~$+$~jet detection is presented in upper and lower panels of Fig.\tref{fig:C0}, respectively.
For our choice of kinematic cuts, values of $C_0$ are almost everywhere higher than 1 nb, thus leading to a quite favorable statistics.
The falloff of both LLA and NLA predictions when $\DY$ grows has been already observed in other semi-hard reactions featuring forward/backward two-particle final states, such as di-jet\tcite{Celiberto:2015yba}, Higgs-jet\tcite{Celiberto:2020tmb,Celiberto:2021fjf} and so on. It comes out as the net combination of two distinct effects. On one side, the partonic cross section increases with energy, as predicted by BFKL evolution. On the other side, collinear parton distributions dampen the hadronic cross section when $\DY$ becomes larger and larger.

We observe that at BLM scales (right panels of Fig.\tref{fig:C0}) NLA bands are almost entirely nested inside LLA ones, while at natural scales (left panels) they decouple from each other in the large $\DY$-range. The decoupling effect is due to the fact that NLA series are very stable under scale variation, this making the corresponding bands thinner than the LLA ones. Conversely, when $\DY$ increases, these latter shrink in the double $H_b$ channel, while they widen in the $H_b$~$+$~jet one. The peculiar behavior of NLA predictions for $C_0$ will also translate in an increased stability of the azimuthal $R_{n0}$ ratios, as pointed out in Section\tref{ssec:Rnm}.
At variance with light-hadron species emissions (protons, pions and kaons\tcite{Celiberto:2017ptm,Bolognino:2018oth,Celiberto:2020wpk}), where cross sections lose one or more orders of magnitude when passing from natural scales to BLM ones, results for $b$-hadrons are much more stable. This effect, already observed in the double $\Lambda_c$ channel\tcite{Celiberto:2021dzy}, here holds also in the $H_b$~$+$~jet one.

All these features brace the message that a stability of our $\DY$-distributions is reached via heavy-flavor emissions, and it becomes strongly manifest when bottom-flavored bound states are detected.


\subsection{Azimuthal correlations}
\label{ssec:Rnm}

Analogously to $C_0$ (see Eq.\eref{DY_distribution}), we define the phase-space integrated higher azimuthal coefficients, $C_{n > 0}$. Thus, we can study their ratios
\begin{widetext}
\begin{equation}
 \label{Rnm}
 R_{nm} \equiv \frac{C_n}{C_m} =
 \frac{
 \int_{y_1^{\rm min}}^{y_1^{\rm max}} \drv y_1
 \int_{y_2^{\rm min}}^{y_2^{\rm max}} \drv y_2
 \int_{p_1^{\rm min}}^{p_1^{\rm max}} \drv |\vec p_1|
 \int_{p_2^{\rm min}}^{p_2^{\rm max}} \drv |\vec p_2|
 \, \,
 \delta (\DY - (y_1 - y_2))
 \, \,
 {\cal C}_n
 }
 {
 \int_{y_1^{\rm min}}^{y_1^{\rm max}} \drv y_1
 \int_{y_2^{\rm min}}^{y_2^{\rm max}} \drv y_2
 \int_{p_1^{\rm min}}^{p_1^{\rm max}} \drv |\vec p_1|
 \int_{p_2^{\rm min}}^{p_2^{\rm max}} \drv |\vec p_2|
 \, \,
 \delta (\DY - (y_1 - y_2))
 \, \,
 {\cal C}_m
 }
 \, .
\end{equation}
\end{widetext}
The $R_{n0}$ ratios have a straightforward physical interpretation, being the azimuthal-correlation moments $\langle \cos n \varphi \rangle$, while the ones without zero indices represent ratios of correlations, that were originally proposed in Refs.~\cite{Vera:2006un,Vera:2007kn}. We investigate the behavior of the $R_{nm}$ moments as functions of $\DY$ and in the kinematic ranges defined in Section\tref{ssec:C0}.

\begin{figure*}[b]
\centering

\includegraphics[scale=0.53,clip]{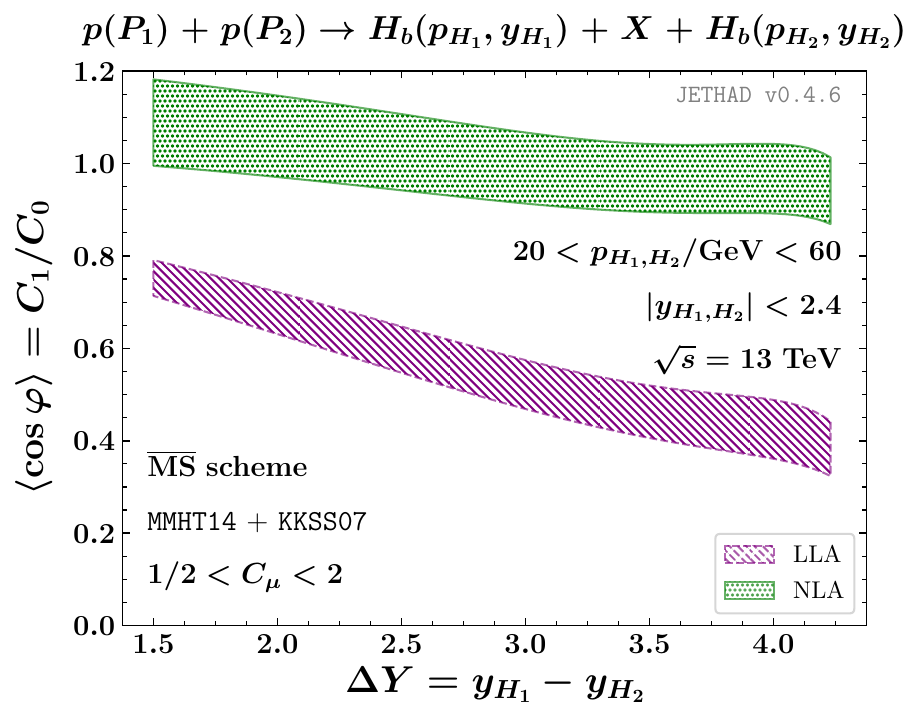}
\includegraphics[scale=0.53,clip]{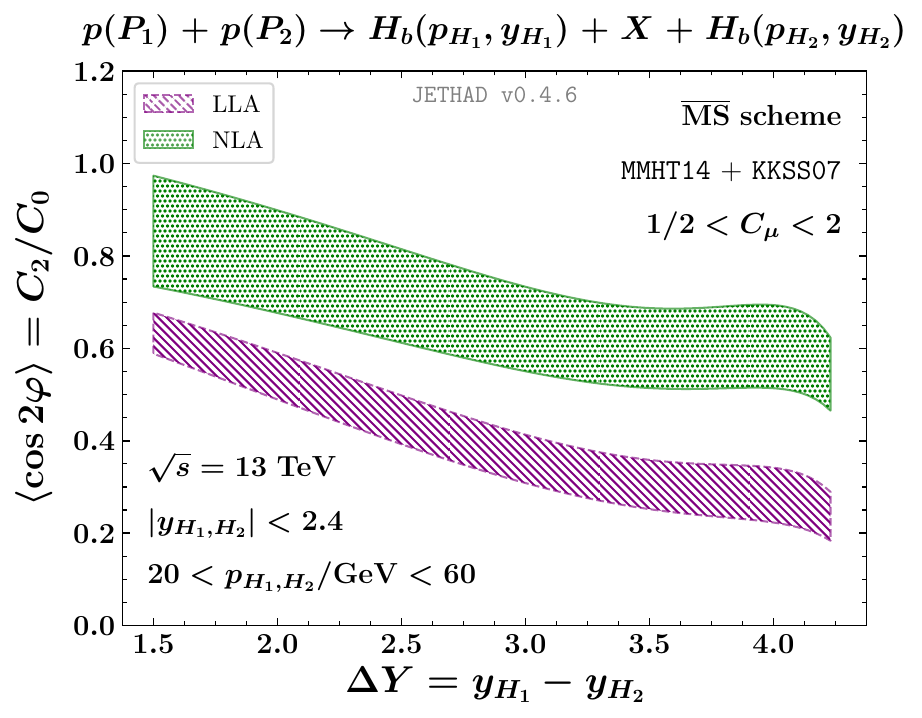}

\includegraphics[scale=0.53,clip]{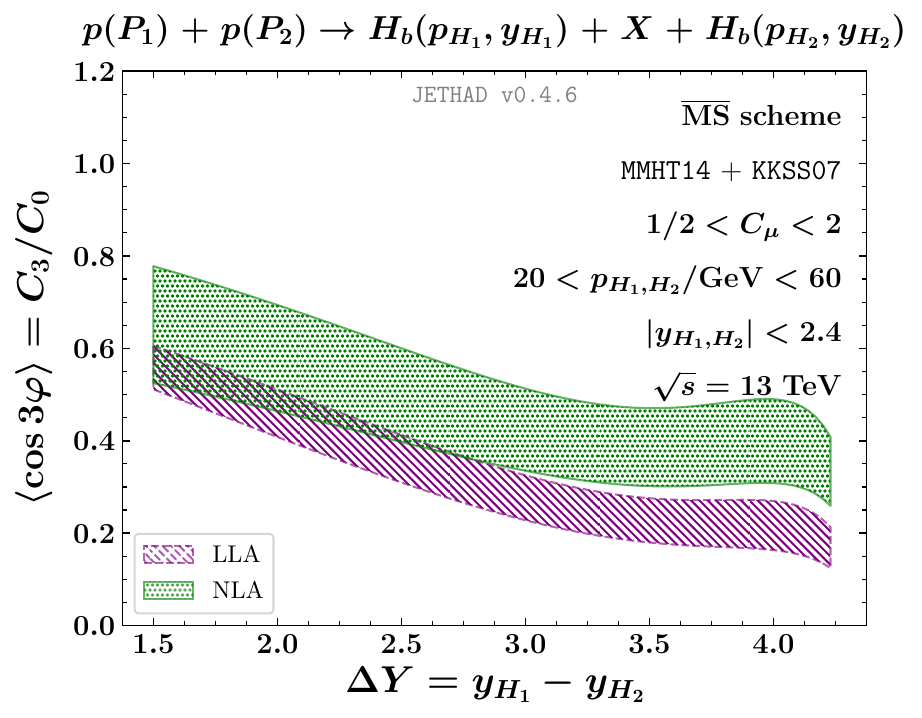}
\includegraphics[scale=0.53,clip]{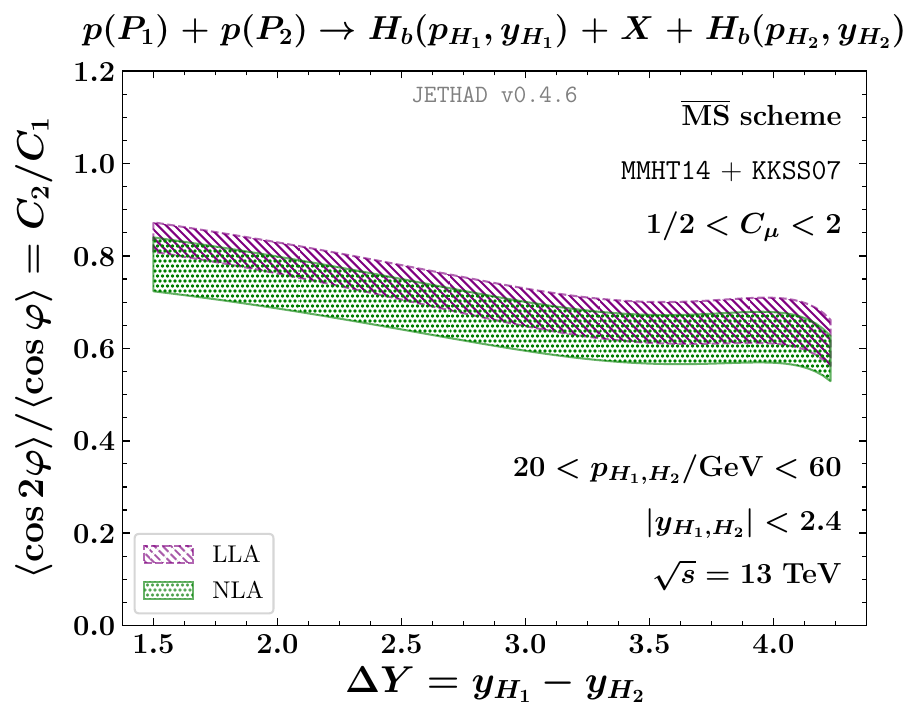}

\caption{$\DY$-shape of azimuthal correlations, $R_{nm} \equiv C_{n}/C_{m}$, in the double $H_b$ channel, at natural scales, and for $\sqrt{s} = 13$ TeV. Text boxes inside panels show transverse-momentum and rapidity ranges. Uncertainty bands embody the combined effect of scale variation and phase-space multi-dimensional integration.}
\label{fig:Rnm_HbHb_NS}
\end{figure*}

\begin{figure*}[b]
\centering

\includegraphics[scale=0.53,clip]{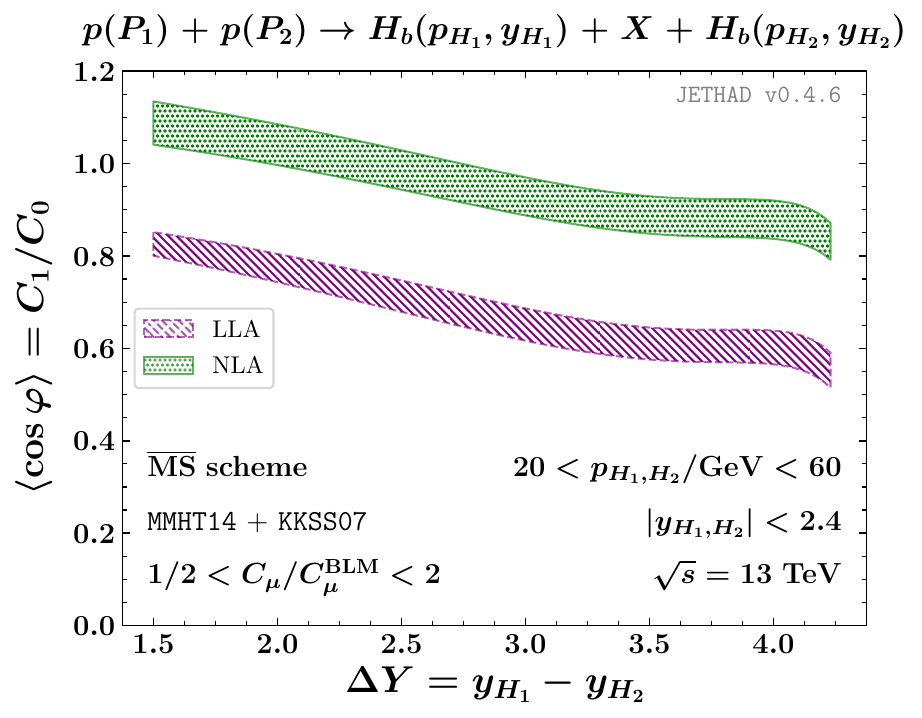}
\includegraphics[scale=0.53,clip]{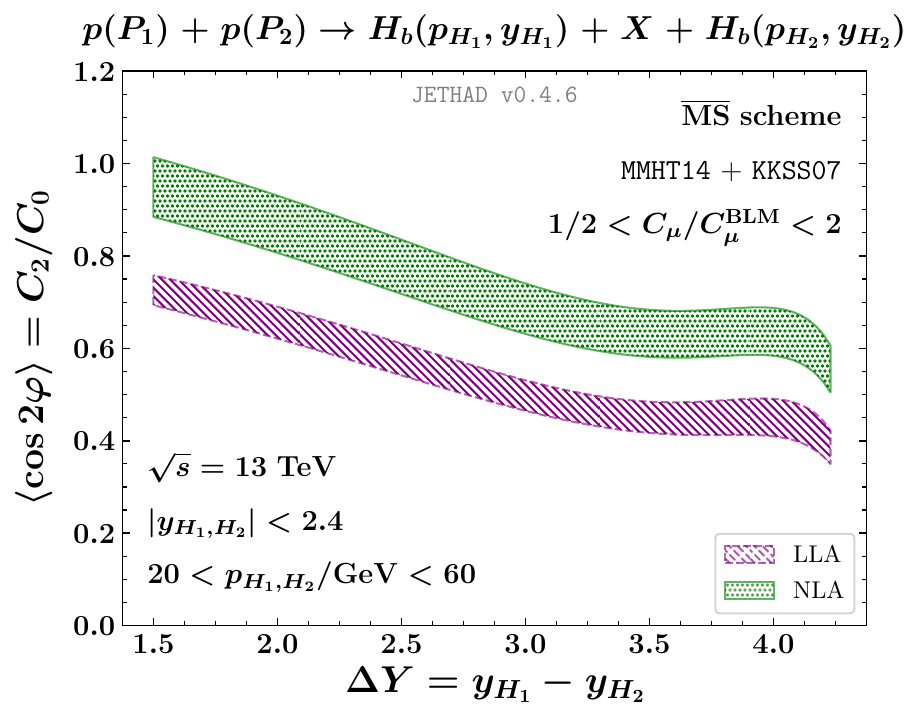}

\includegraphics[scale=0.53,clip]{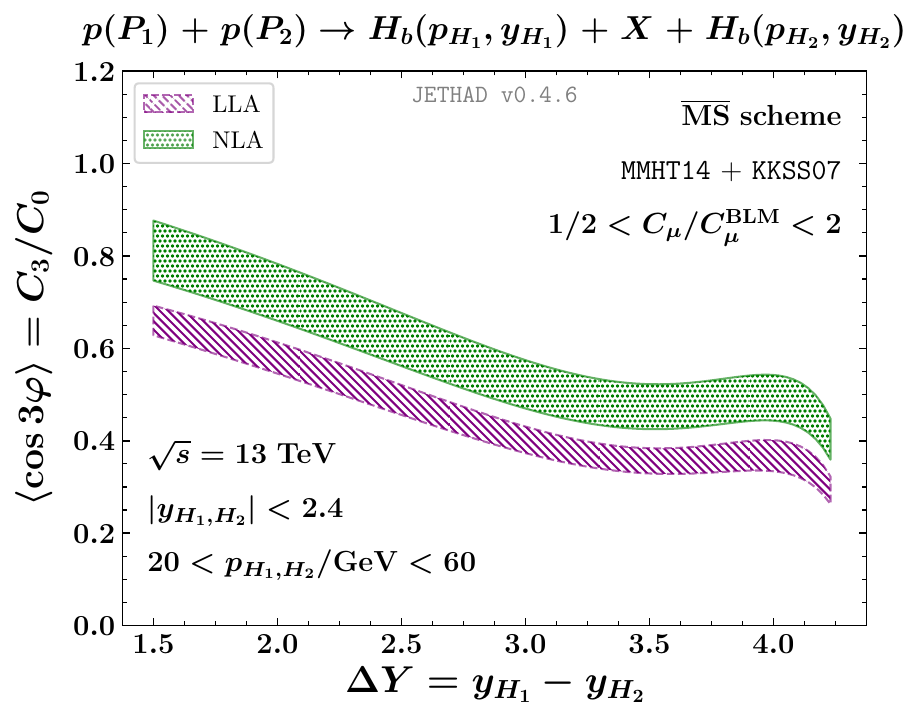}
\includegraphics[scale=0.53,clip]{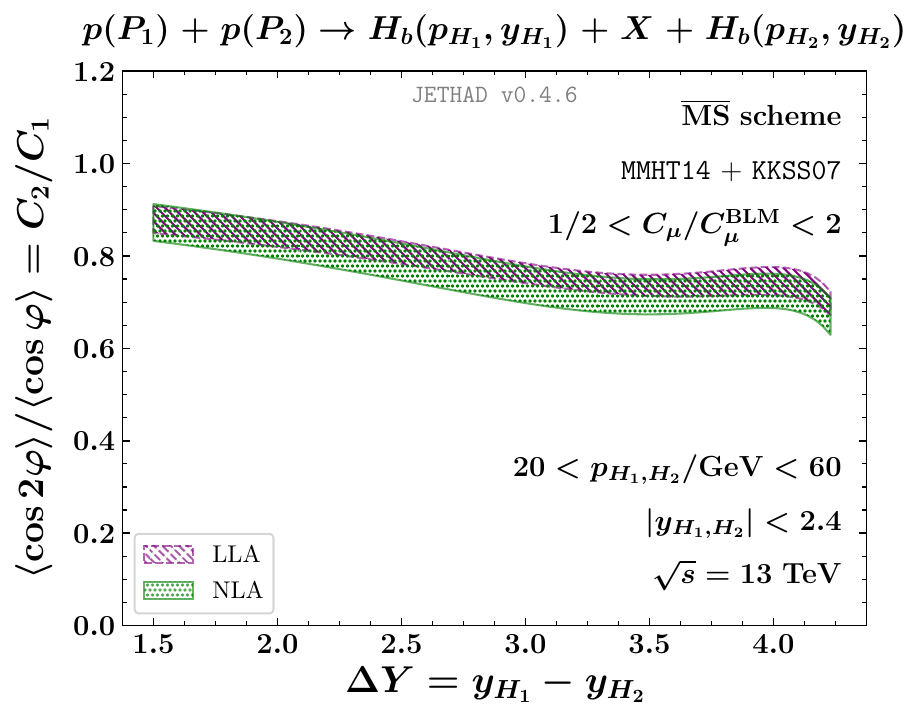}

\caption{$\DY$-shape of azimuthal correlations, $R_{nm} \equiv C_{n}/C_{m}$, in the double $H_b$ channel, at BLM scales, and for $\sqrt{s} = 13$ TeV. Text boxes inside panels show transverse-momentum and rapidity ranges. Uncertainty bands embody the combined effect of scale variation and phase-space multi-dimensional integration.}
\label{fig:Rnm_HbHb_BLM}
\end{figure*}

\begin{figure*}[b]
\centering
\includegraphics[scale=0.53,clip]{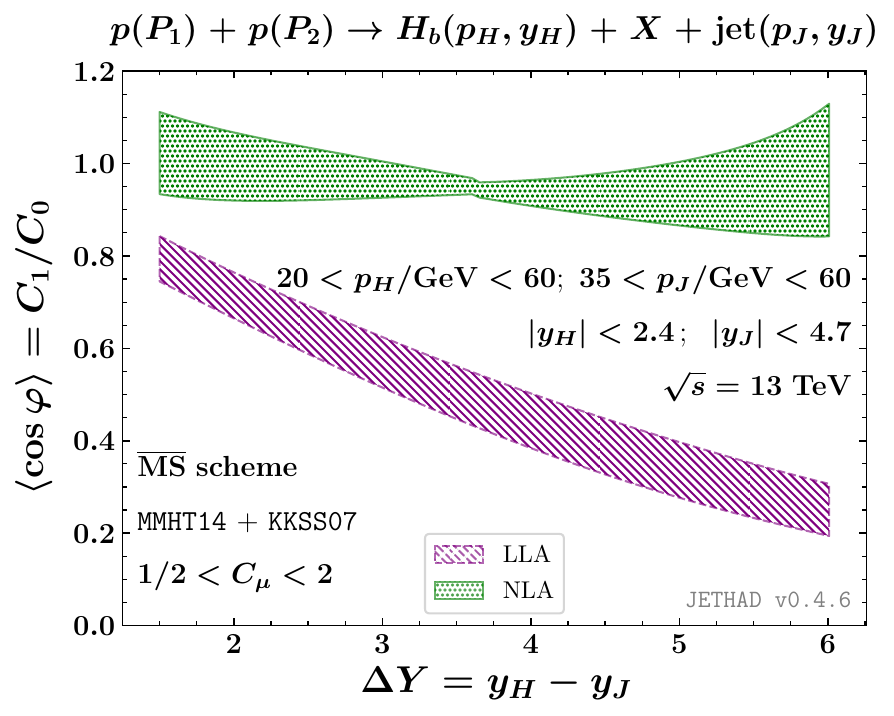}
\includegraphics[scale=0.53,clip]{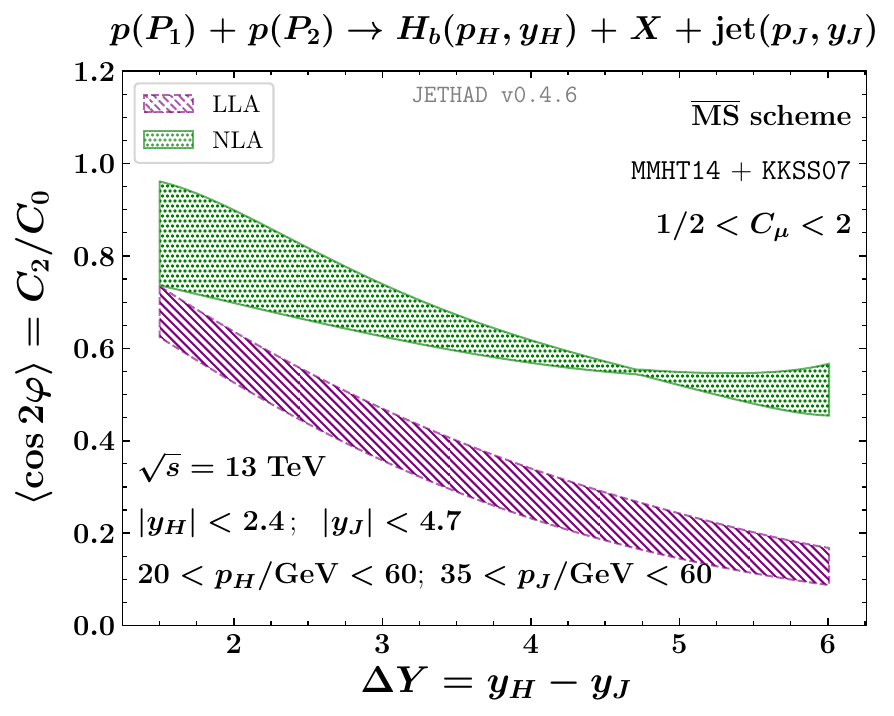}

\includegraphics[scale=0.53,clip]{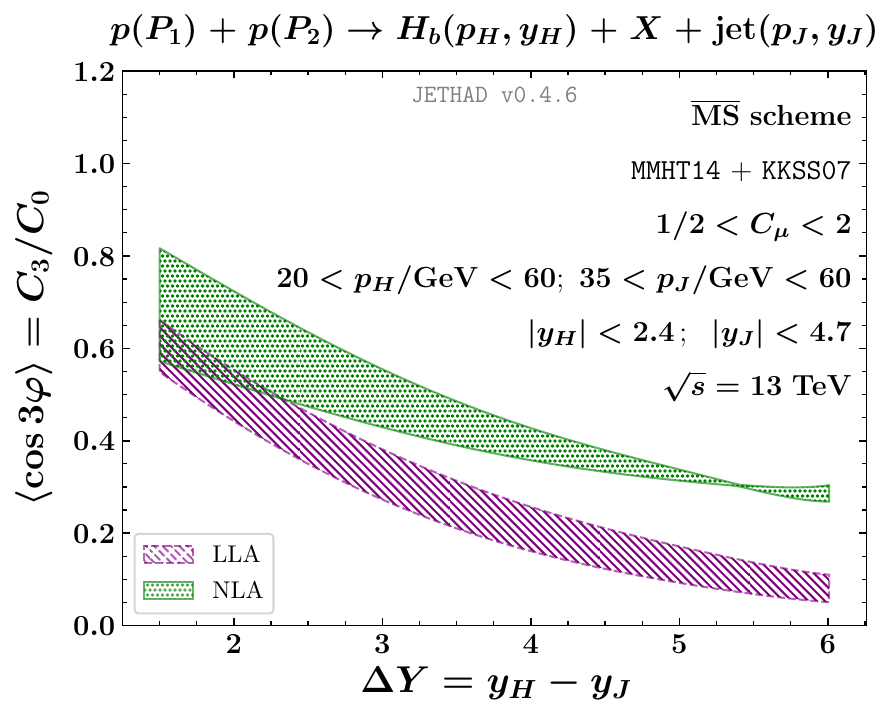}
\includegraphics[scale=0.53,clip]{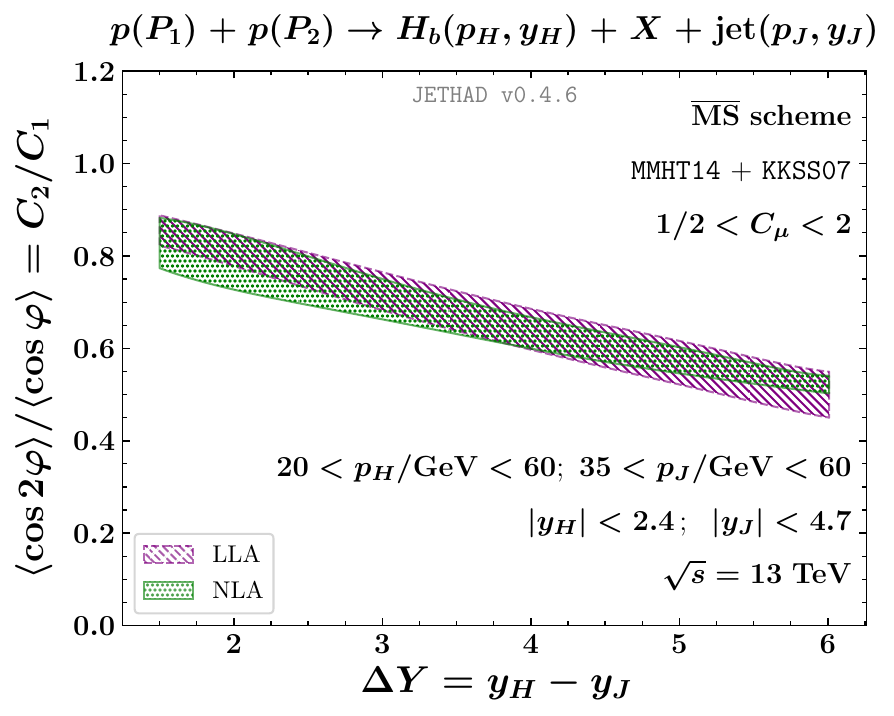}
\caption{$\DY$-shape of azimuthal correlations, $R_{nm} \equiv C_{n}/C_{m}$, in the $H_b$~$+$~jet channel, at natural scales, and for $\sqrt{s} = 13$ TeV. Text boxes inside panels show transverse-momentum and rapidity ranges. Uncertainty bands embody the combined effect of scale variation and phase-space multi-dimensional integration.}
\label{fig:Rnm_HbJ_NS}
\end{figure*}

\begin{figure*}[b]
\centering
\includegraphics[scale=0.53,clip]{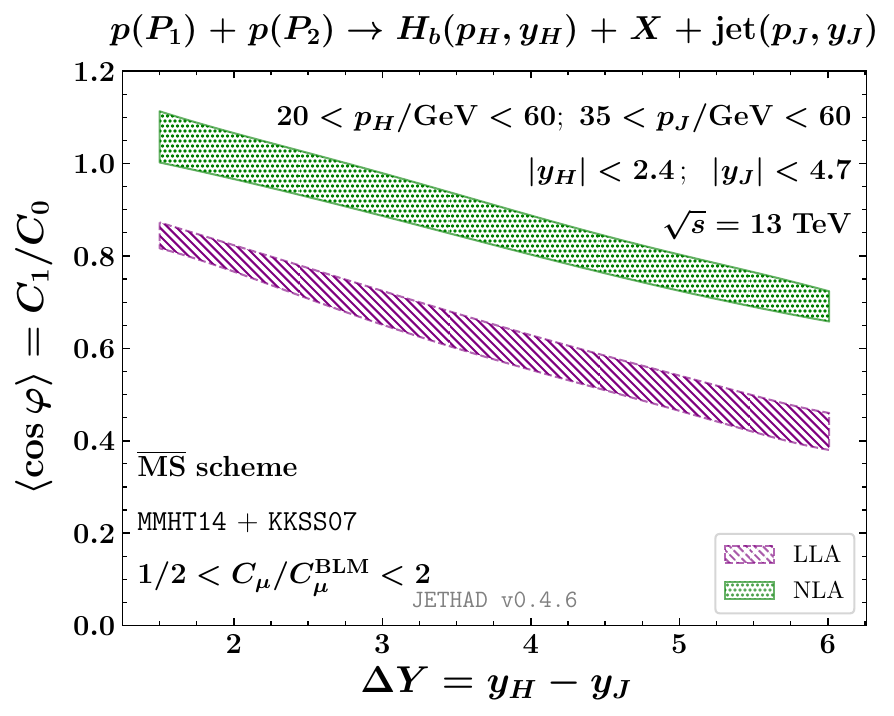}
\includegraphics[scale=0.53,clip]{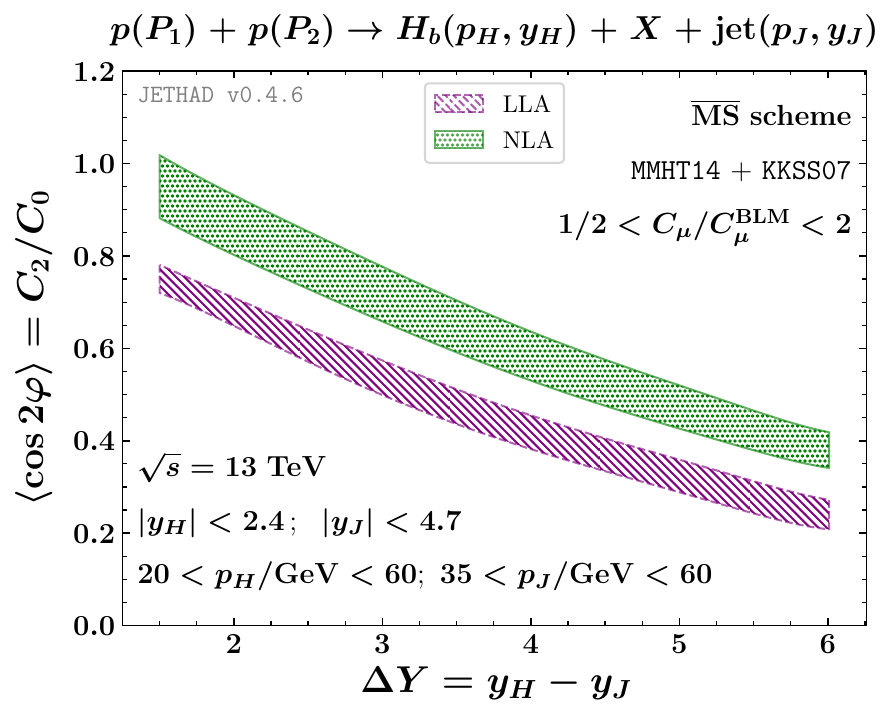}

\includegraphics[scale=0.53,clip]{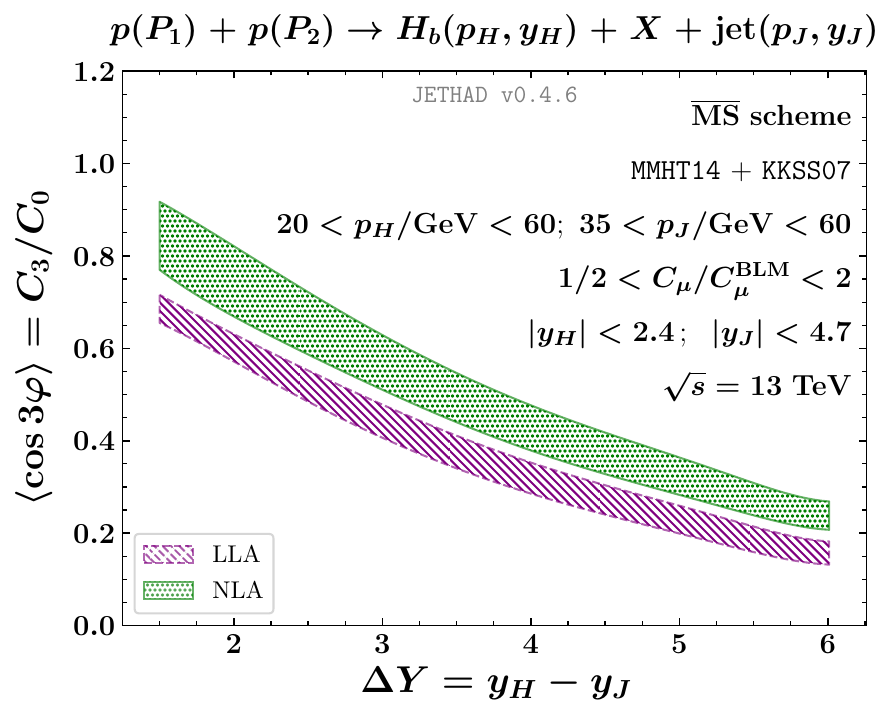}
\includegraphics[scale=0.53,clip]{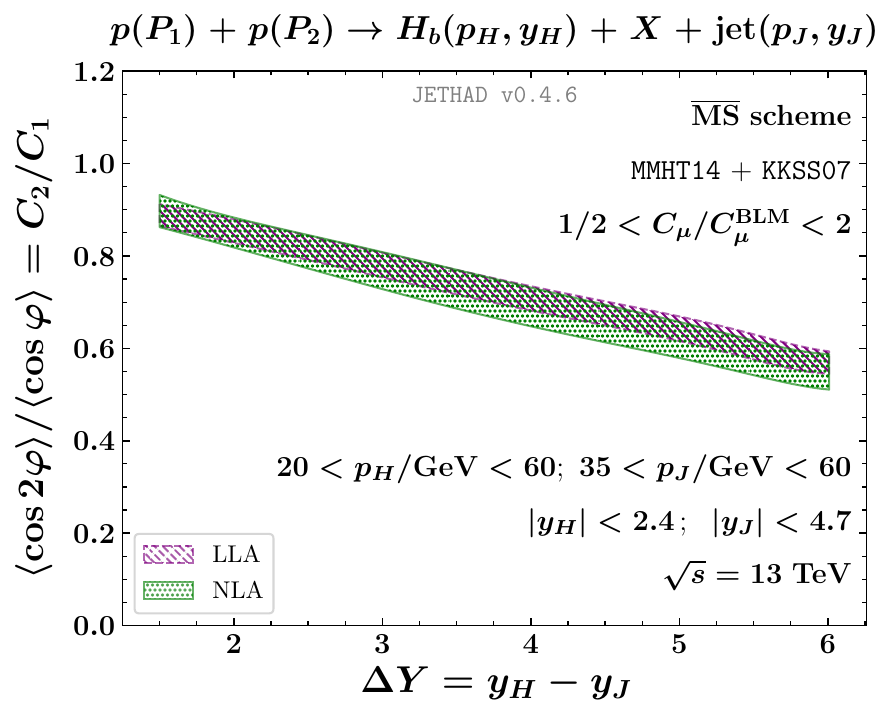}
\caption{$\DY$-shape of azimuthal correlations, $R_{nm} \equiv C_{n}/C_{m}$, in the $H_b$~$+$~jet channel, at BLM scales, and for $\sqrt{s} = 13$ TeV. Text boxes inside panels show transverse-momentum and rapidity ranges. Uncertainty bands embody the combined effect of scale variation and phase-space multi-dimensional integration.}
\label{fig:Rnm_HbJ_BLM}
\end{figure*}

We present results for the double $H_b$ channel at natural and at BLM scales in Figs.\tref{fig:Rnm_HbHb_NS} and\tref{fig:Rnm_HbHb_BLM}, respectively. From a first inspection of our plots we fairly note that the onset of high-energy dynamics has come into play.
All ratios decrease when $\DY$ grows, since the weight of undetected gluons becomes more and more relevant, as predicted by BFKL. This leads to a decorrelation pattern in the azimuthal plane, which is stronger at LLA.
We observe that predictions at natural scales are close in shape to the corresponding BLM-optimized ones. NLA bands are thicker at natural scales, but still thinner than what one finds for double $\Lambda_c$ emissions. As anticipated in Section\tref{ssec:C0}, this increased stability is mostly due, for $R_{n0}$ correlations, to the relatively small uncertainty on $C_0$ that propagates in the ratio.

The $\DY$-behavior of azimuthal ratios in the $H_b$~$+$~jet channel at natural scales is shown in Figs.\tref{fig:Rnm_HbJ_NS}. Here the NLA $R_{n0}$ correlations exhibit a strong sensitivity to scale variation. More in particular, the upper bound of uncertainty bands, given by predictions for $C_\mu=2$, at some point on $\DY$-axis crosses the lower bound, determined by results for $C_\mu=1/2$. The crossing point depends on the value of $n$ and goes from $\DY\gtrsim3.6$ for $n=0$ to $\DY\gtrsim5.5$ for $n=2$, while it is not present for the $R_{21}$ ratio, where LLA and NLA bands almost overlap.
Conversely, corresponding results for $R_{nm}$ moments at BLM scales (see Fig.\tref{fig:Rnm_HbJ_BLM}) present a $\DY$-shape similar to predictions for the double $H_b$ emission.
Although the sensitivity on scale variation is strong, the possibility of performing analyses at natural scales when jet emissions are considered is itself a signal of a partial stability reached by our azimuthal correlations.
Indeed, such a result cannot be obtained in other semi-hard reaction studied at NLA, as Mueller--Navelet dijet or ligther-hadron~$+$~jet production, where instabilities emerging at natural scales are so strong to prevent any realistic analysis. Future studies, postponed to the medium-term future, are needed to unveil the connection between the sensitivity of  $R_{n0}$ ratio on scale variation and other potential sources of uncertainty, as the jet algorithm selection (see Section\tref{ssec:ingredients}).

As a general remark, we note that the value of $R_{10}$ exceeds one for small $\DY$-values. This is an unphysical effect generated by terms, power-suppressed in energy and missed by the BFKL resummation, that start to become relevant in the low $\DY$-range, thus calling for a treatment beyond the scope of this paper.

\subsection{Double differential $p_T$-distribution}
\label{ssec:pT}

Cross sections and azimuthal-angle correlations differential in the final-state rapidity interval, $\DY$, are excellent testing grounds for the high-energy resummation.
However, in order to probe regimes where other resummation dynamics are also relevant, more differential distributions in the $p_T$-spectrum are needed. Indeed, when the measured transverse momenta range in wider windows, other regions that are contiguous to the the strict semi-hard one get probed. 

On one hand, when the transverse momenta are very large or their mutual distance is large, the weight of DGLAP-type logarithms as well as \emph{threshold} contaminations\tcite{Bonciani:2003nt,deFlorian:2005fzc,Muselli:2017bad} grows, thus making the description by our formalism inadequate.
On the other hand, in the very low-$p_T$ limit a pure high-energy treatment would also fail since large transverse-momentum logarithms entering the perturbative series are systematically neglected by BFKL. Moreover, \emph{diffusion-pattern} effects~\cite{Bartels:1993du} (see also Refs.~\cite{Caporale:2013bva,Ross:2016zwl}) would become more and more relevant up to spoiling the convergence of the high-energy series.
The most effective way to account for those $p_T$-logarithms is performing an all-order transverse-momentum (TM) resummation (see Refs.\tcite{Catani:2000vq,Bozzi:2005wk,Bozzi:2008bb,Catani:2010pd,Catani:2011kr,Catani:2013tia,Catani:2015vma} and references therein).

Recently, TM-resummed predictions were proposed for the hadroproduction of inclusive paired systems, as photon\tcite{Cieri:2015rqa,Becher:2020ugp,Neumann:2021zkb} and Higgs\tcite{Ferrera:2016prr} pairs.
The first joint resummation of TM logarithms coming from the emission of two distinct particles was considered in Ref.\tcite{Monni:2019yyr}, where the concurrent measurement of the Higgs and the leading-jet transverse momenta in hadronic Higgs-boson emissions was studied up to the next-to-next-to-leading-logarithmic order via the {\tt RadISH} code\tcite{Bizon:2017rah}. Those studies were then extended to TM-resummed differential observables for color-singlet channels, as the fully leptonic $W^+W^-$ production at the LHC\tcite{Kallweit:2020gva}.
The double-differential spectrum on transverse momentum and azimuthal angle for weak gauge-boson production ($W^\pm$ or $Z^0$) was recently investigated in the TM context via a soft-collinear effective theory approach\tcite{Ju:2021lah}.

An additional issue arises when heavy-flavored emissions are considered. In our case, when the $p_T$ of a $b$-flavored hadron is very small, energy scales are close to the DGLAP-evolution threshold given by the $b$-quark mass, even crossing it when fractions of natural scales are selected (\emph{e.g.}, for $C_\mu=1/2$). Thus the validity of a VFNS treatment, upon which PDFs and FFs employed in this work are built, does not hold anymore. Here a more sophisticated description based on the GM-VFNS needs to be accounted for.

In this Section we study distributions at fixed $\DY$-values and differential in the transverse momenta of both the emitted particles, in the range 10~GeV~$< |\vec p_{1,2}| <$~100~GeV, where energy logarithms rising from the semi-hard scale ordering are relevant, but at the same time also contaminations coming from $p_T$-logarithms are expected. We propose this analysis without pretension of catching all the dominant features of these observables by the hand of our hybrid factorization, but rather to set the ground for futures studies where the interplay of different resummations (among all BFKL, TM, and threshold one) is deeply investigated.

We build the transverse-momentum double differential cross section as
\begin{widetext}
\begin{equation}
\label{pT_distribution}
 \frac{\drv \sigma(|\vec p_{1,2}|, \DY, s)}{\drv |\vec p_1| \drv |\vec p_2| \drv \DY} =
 \int_{y^{\rm min}_1}^{y^{\rm max}_1} \drv y_1
 \int_{y^{\rm min}_2}^{y^{\rm max}_2} \drv y_2
 \; \delta (\DY - (y_1 - y_2))
 \; {\cal C}_0\left(|\vec p_1|, |\vec p_2|, y_1, y_2 \right)
 \, ,
\end{equation}
\end{widetext}
the rapidity ranges of final-state objects being given in Section\tref{ssec:C0}.

\begin{figure*}[b]
\centering

   \includegraphics[scale=0.41,clip]{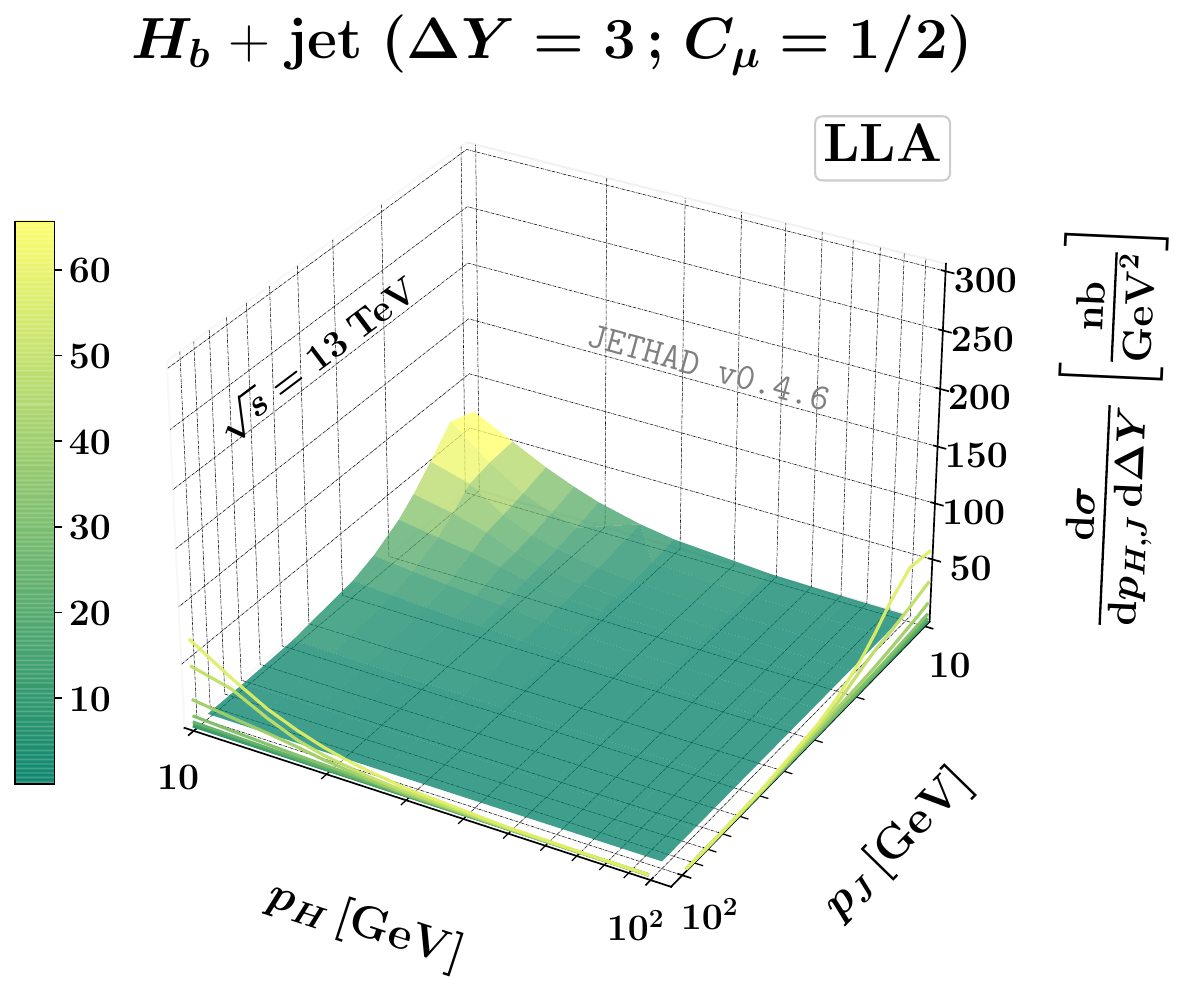}
   \hspace{0.25cm}
   \includegraphics[scale=0.41,clip]{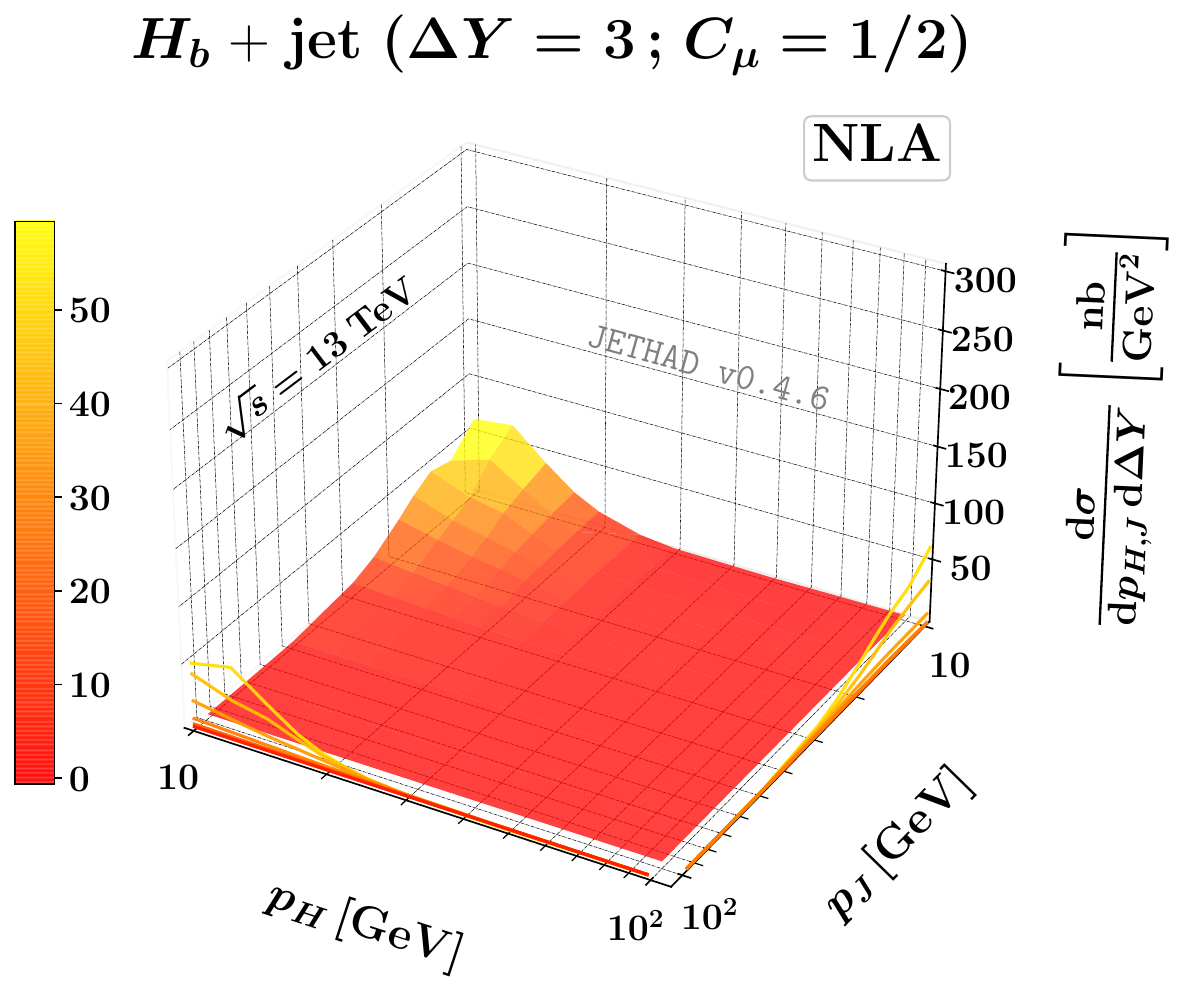}
   \vspace{0.15cm}

   \includegraphics[scale=0.41,clip]{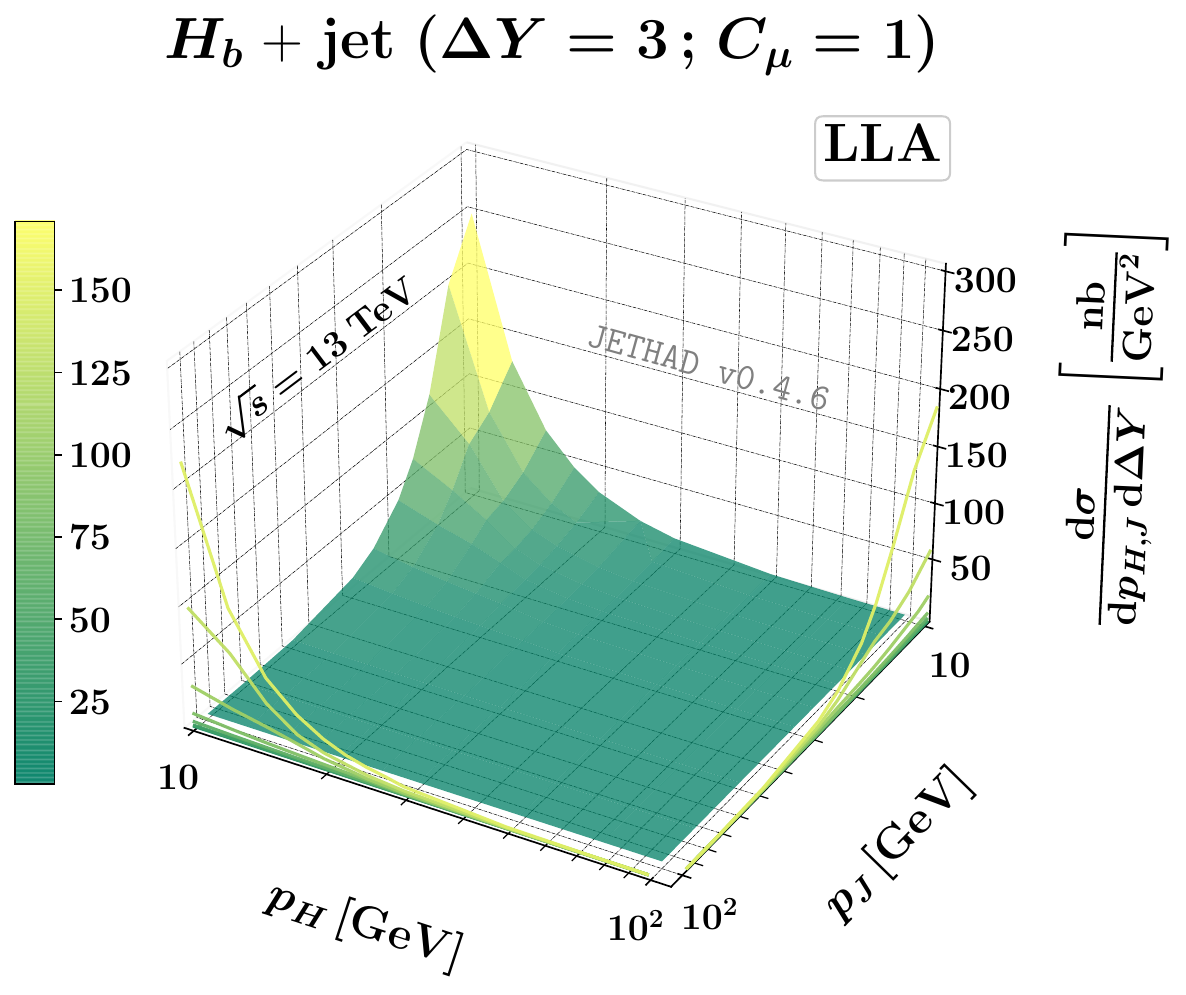}
   \hspace{0.25cm}
   \includegraphics[scale=0.41,clip]{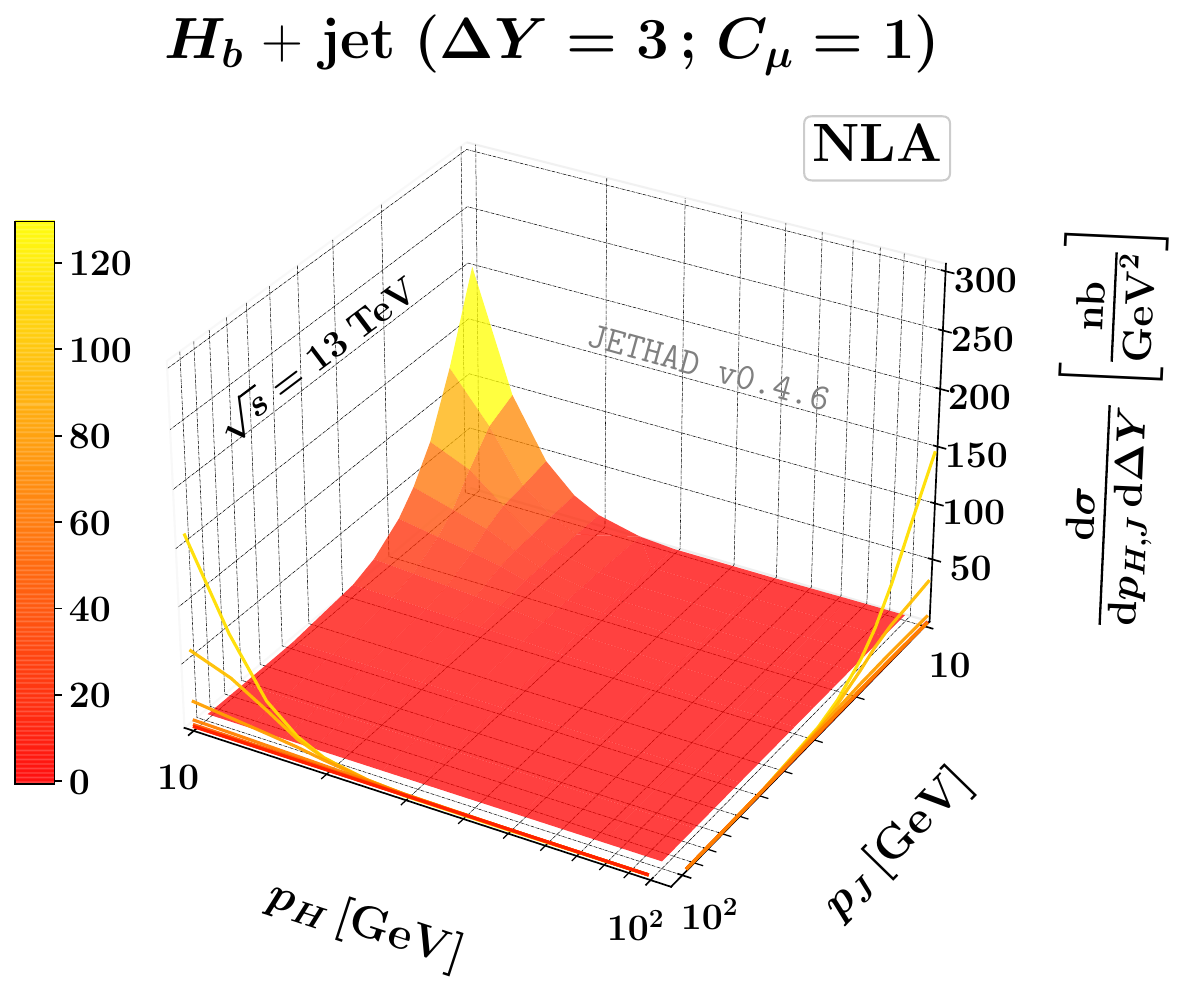}
   \vspace{0.15cm}

   \includegraphics[scale=0.41,clip]{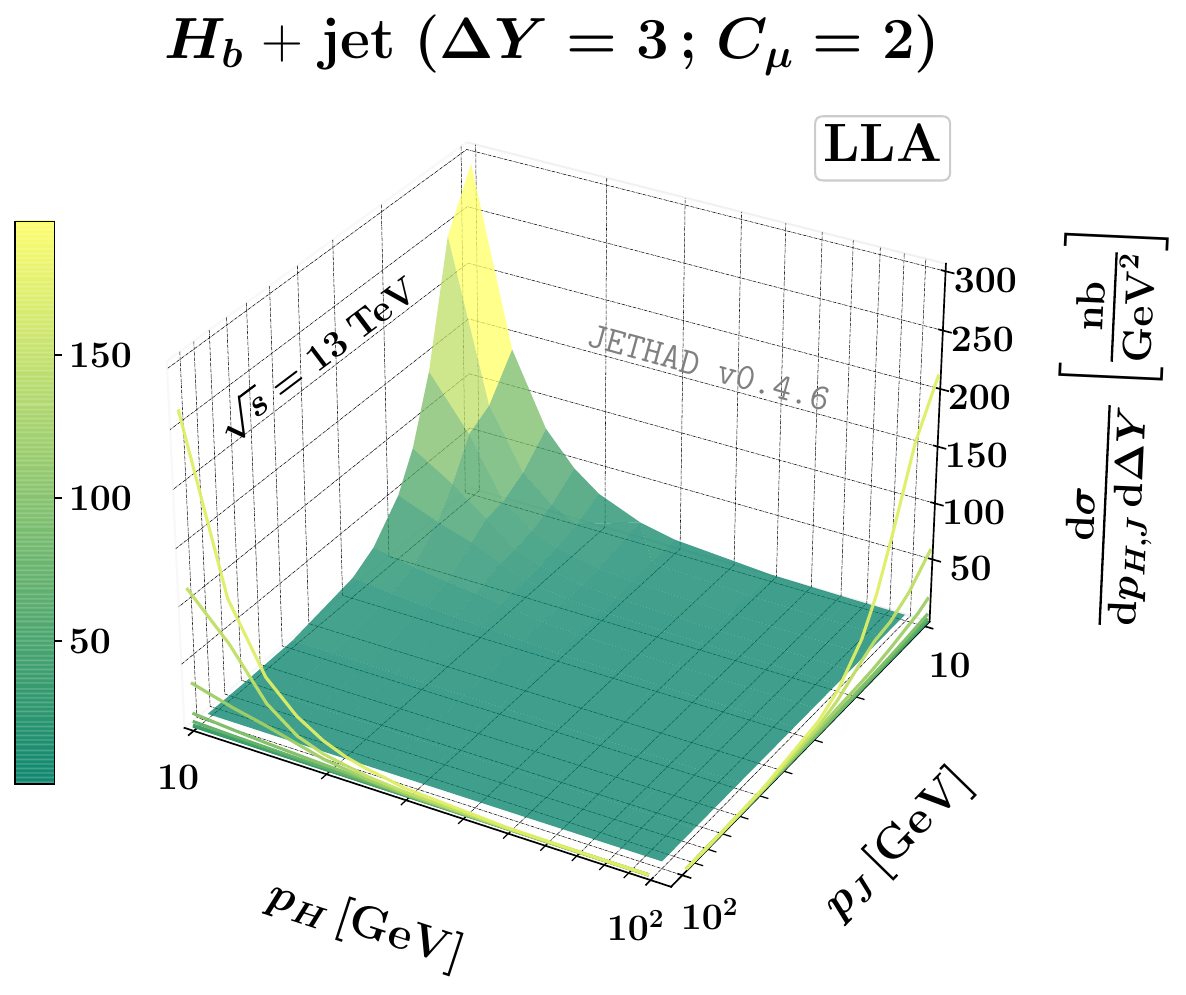}
   \hspace{0.25cm}
   \includegraphics[scale=0.41,clip]{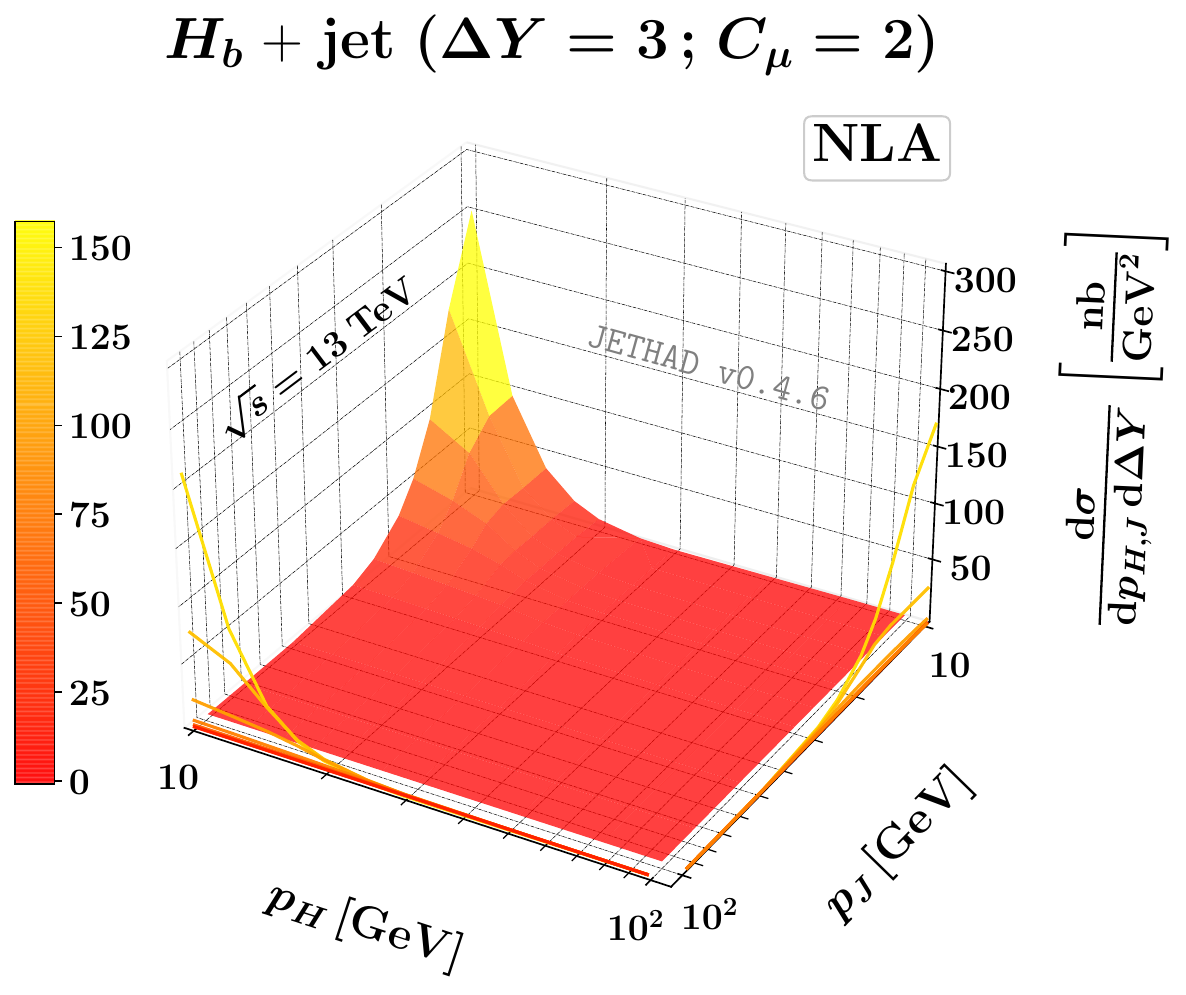}
   \vspace{0.15cm}

\caption{Double differential $p_T$-distribution for the $H_b$~$+$~jet channel at $\DY=3$, $\sqrt{s} = 13$ TeV, and in the LLA (left) and NLA (right) resummation accuracy. Calculations are done at natural scales, and the $C_\mu$ parameter is in the range 1/2 to 2 (from top to bottom).}
\label{fig:Y3-2pT0}
\end{figure*}

\begin{figure*}[b]
\centering

   \includegraphics[scale=0.41,clip]{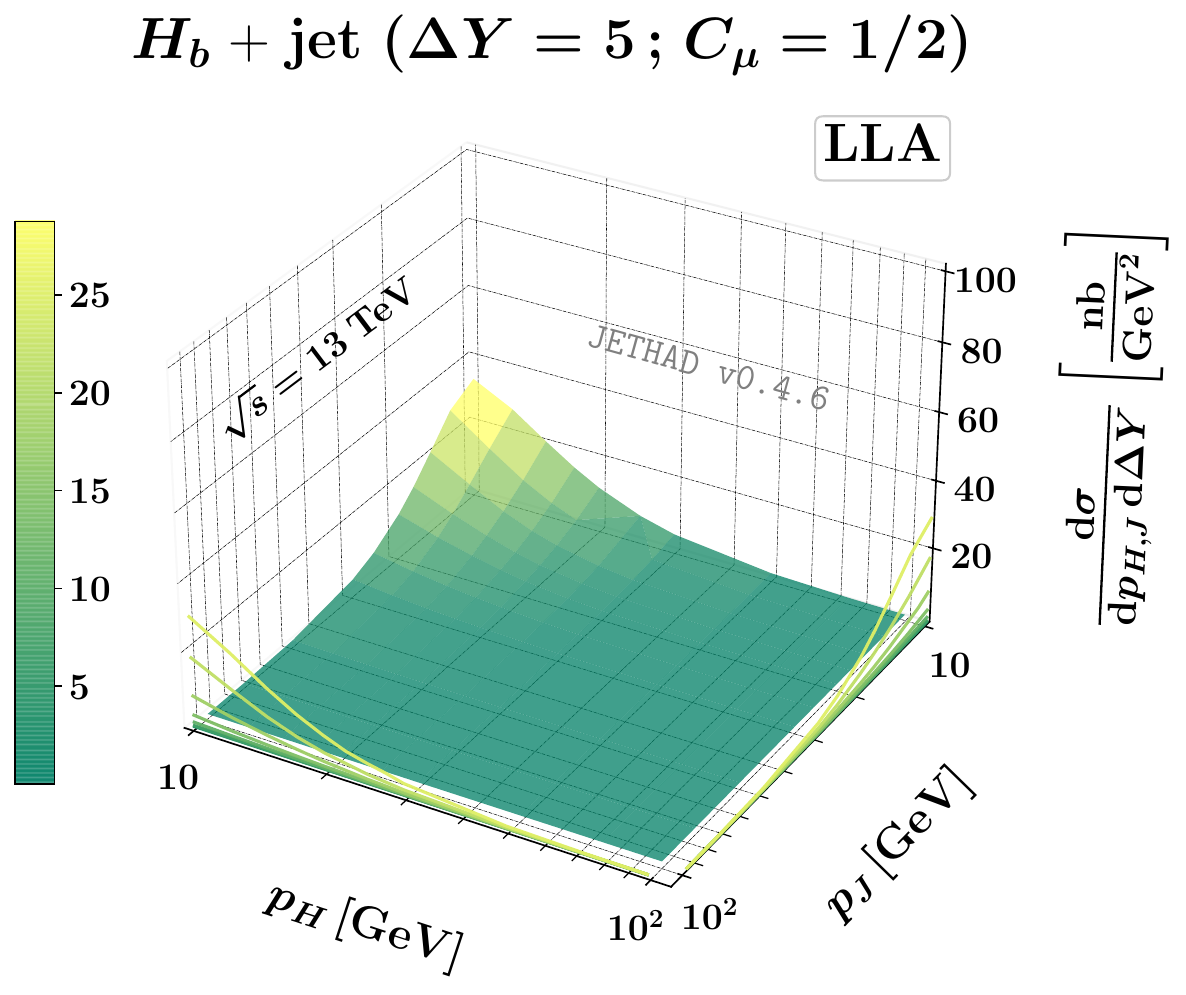}
   \hspace{0.25cm}
   \includegraphics[scale=0.41,clip]{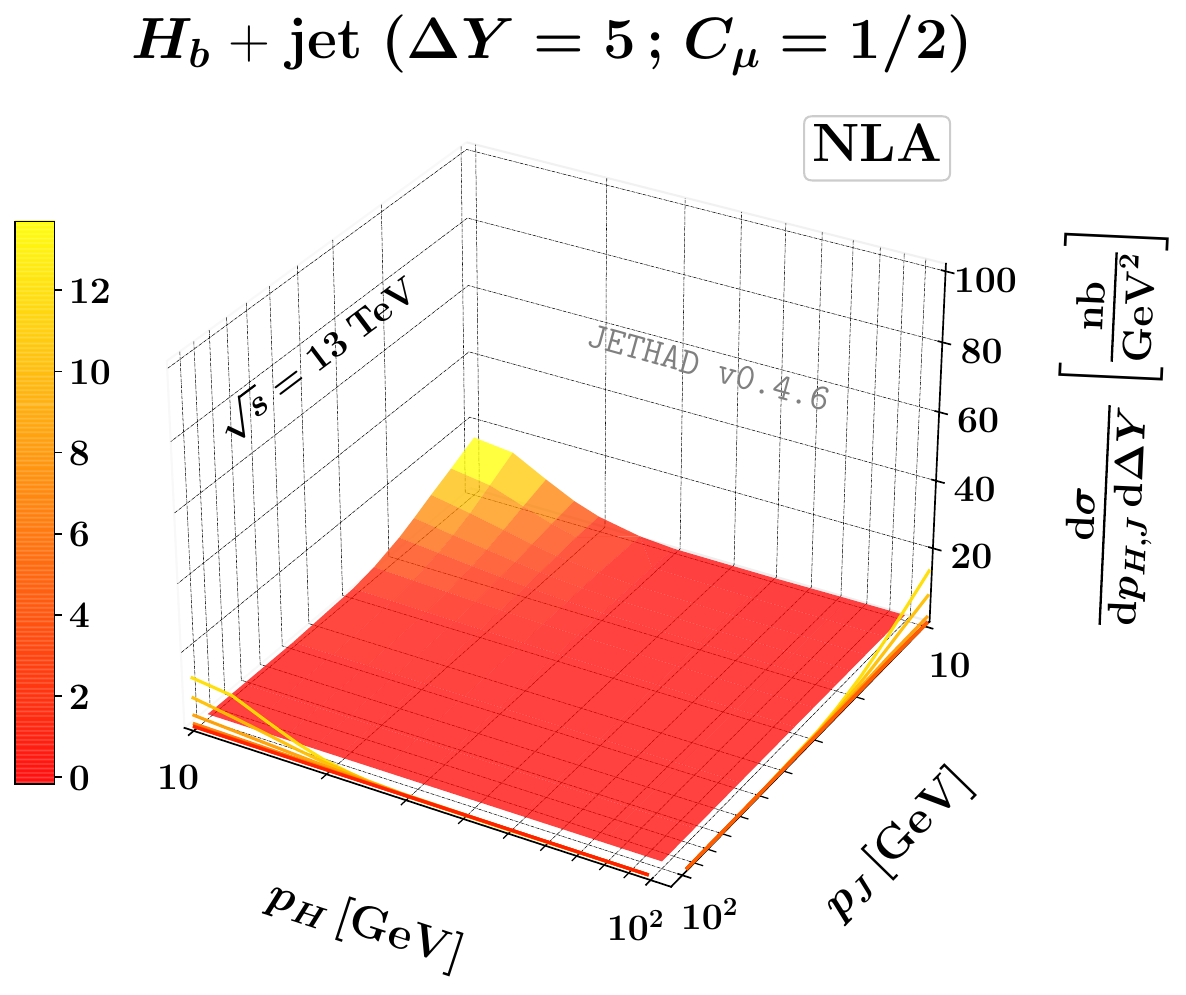}
   \vspace{0.15cm}

   \includegraphics[scale=0.41,clip]{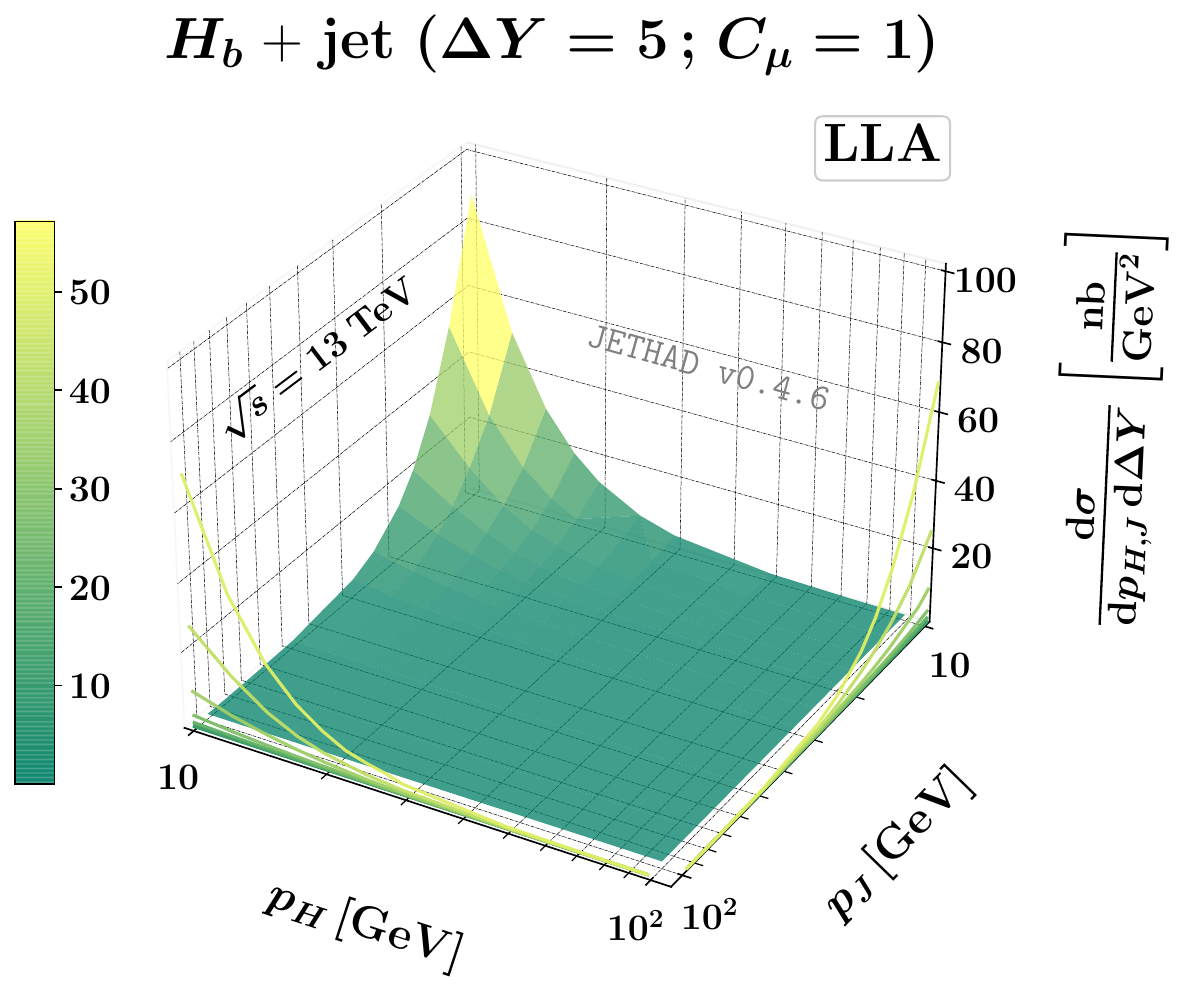}
   \hspace{0.25cm}
   \includegraphics[scale=0.41,clip]{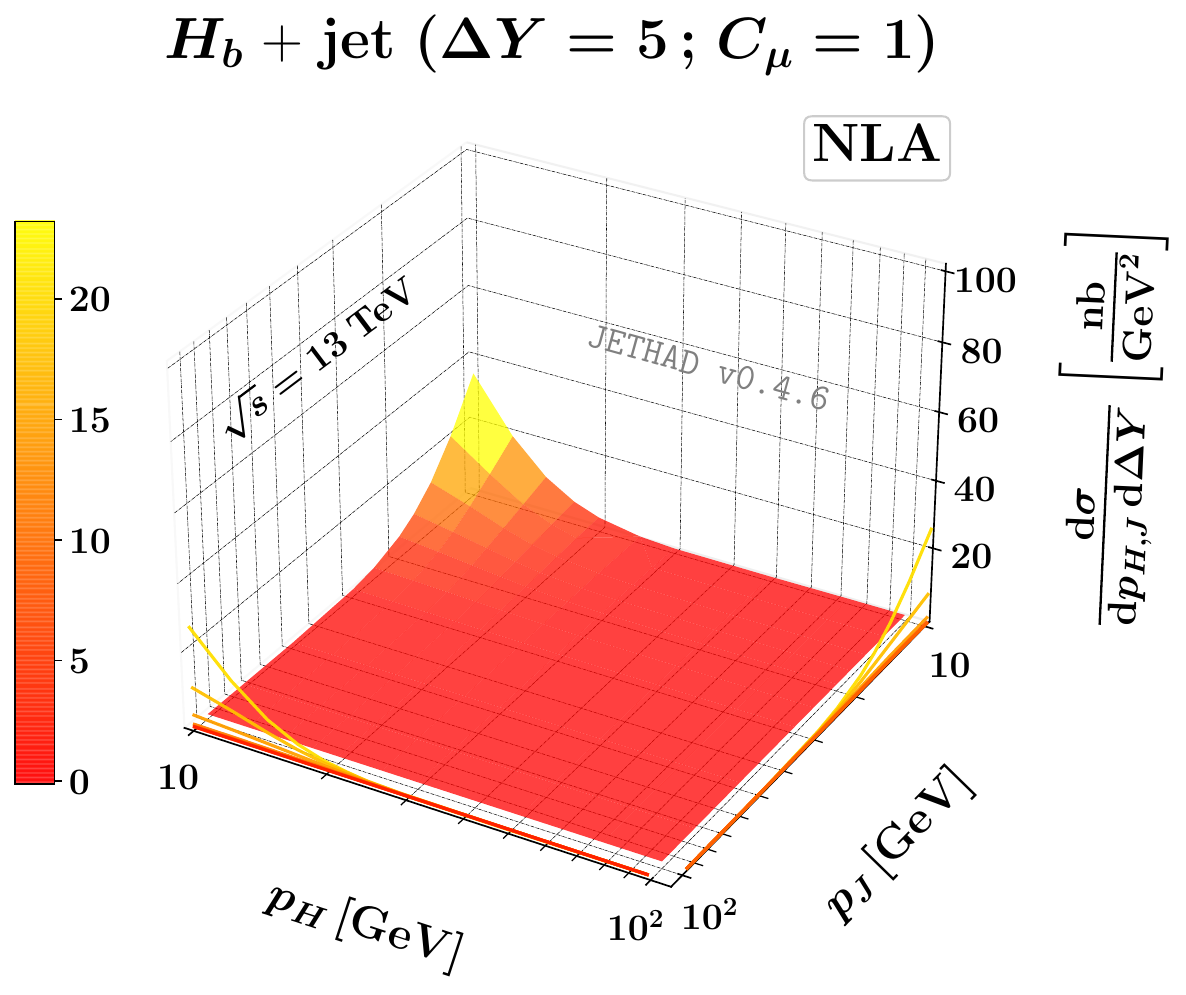}
   \vspace{0.15cm}

   \includegraphics[scale=0.41,clip]{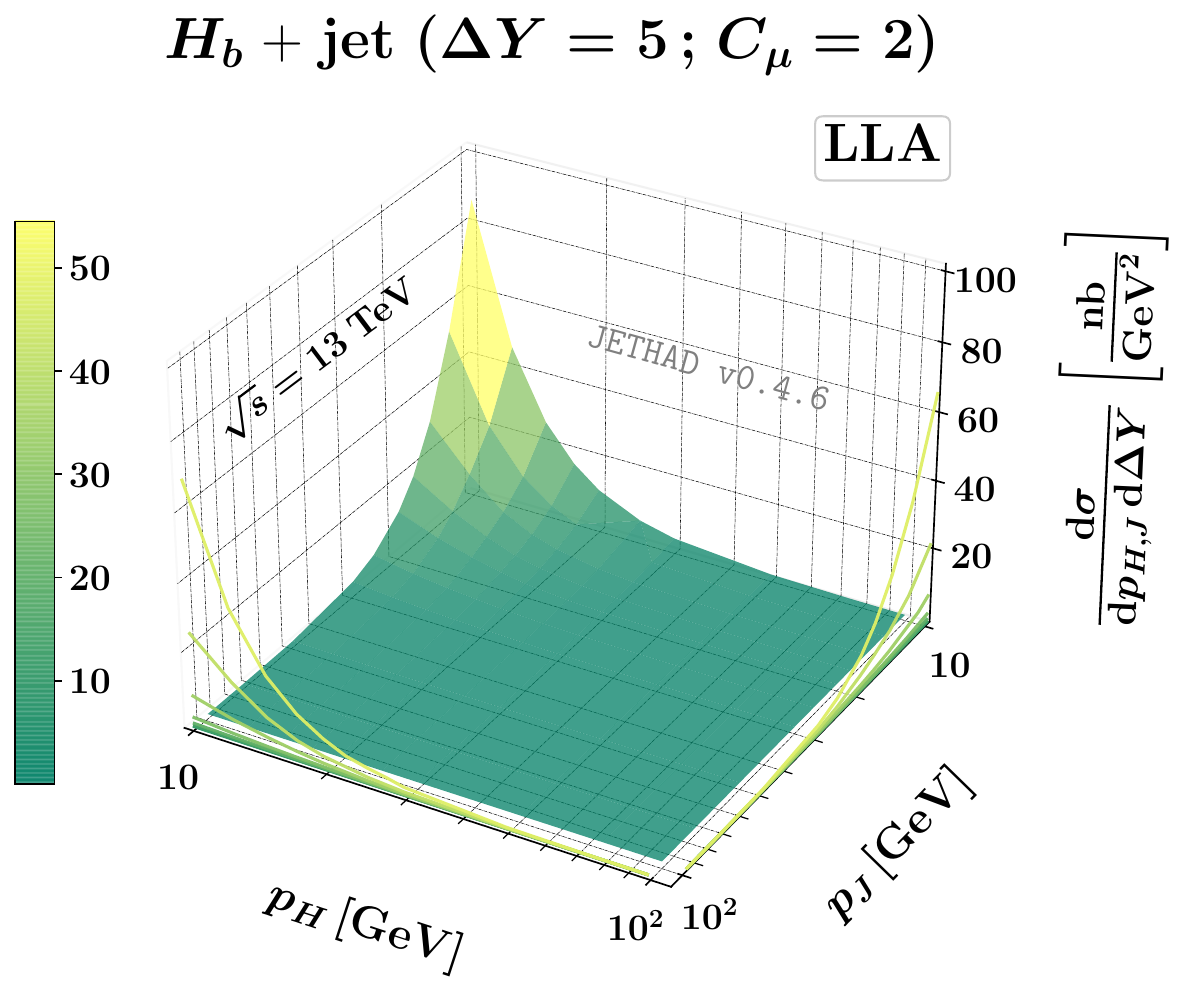}
   \hspace{0.25cm}
   \includegraphics[scale=0.41,clip]{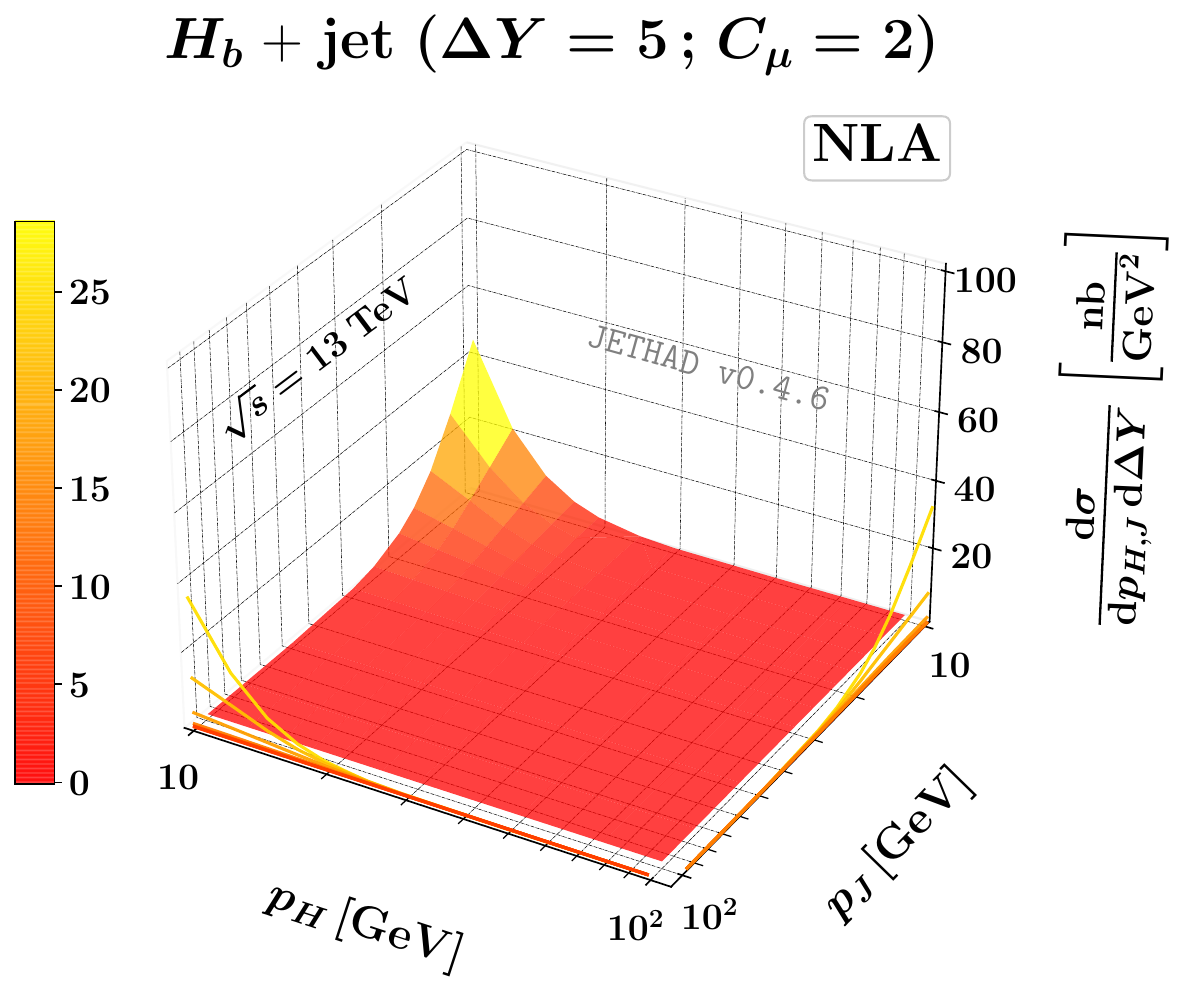}
   \vspace{0.15cm}

\caption{Double differential $p_T$-distribution for the $H_b$~$+$~jet channel at $\DY=5$, $\sqrt{s} = 13$ TeV, and in the LLA (left) and NLA (right) resummation accuracy. Calculations are done at natural scales, and the $C_\mu$ parameter is in the range 1/2 to 2 (from top to bottom).}
\label{fig:Y5-2pT0}
\end{figure*}

Results for our distributions in the $H_b$~$+$~jet channel at $\DY=3$ and $5$ are presented in Figs.\tref{fig:Y3-2pT0} and\tref{fig:Y5-2pT0}, respectively. In this analysis no BLM scale optimization is employed. We note that predictions fall off very fast when the two observable transverse momenta, $|\vec p_H|$ and $|\vec p_J|$, become larger or when their mutual distance grows. As generally predicted by the BFKL dynamics, LLA predictions (left panels) are always larger than NLA ones (right panels). The effect of scale variation (from top to bottom panels) seems to be more relevant with respect to what happens for the $\Delta Y$-distribution and the azimuthal correlations. Furthermore, we do not observe any peak, which could be present in the low-$p_T$ region, namely where TM-resummation effects are dominant and that it is excluded from our analysis.

More quantitative information can be gathered by the inspection of Tables\tref{tab:Y3-2pT0} and\tref{tab:Y5-2pT0}. Here we show numerical values of our distributions for a representative sample of $(|\vec p_H|, |\vec p_J|)$ pairs.
The general trend is that the sensitivity on scale variation of all the predictions grows as we move away from the symmetric $p_T$-region, $|\vec p_H| \simeq |\vec p_J|$. Moreover, for almost all the considered $p_T$-pairs in Tables\tref{tab:Y3-2pT0} and\tref{tab:Y5-2pT0}, LLA results decrease when the $C_\mu$ scale parameter grows. Conversely, NLA results tend to oscillate around $C_\mu = 1$, which seems to act as a critical point for them. This clearly indicates that our distributions are more stable on scale variation when higher-order corrections are included. At the same time, their sensitivity on $C_\mu$ is almost of the same order (up to 45\%) for both LLA and NLA cases when $\Delta Y = 3$~(Table\tref{tab:Y3-2pT0}), while it is roughly halved when passing from LLA (up to 50\%) to NLA (up to 25\%) for $\Delta Y = 5$~(Table\tref{tab:Y5-2pT0}). This reflects the fact that the stabilizing effect of higher-order corrections is more pronounced when we go through the BFKL-sensitive region, \emph{i.e.} when $\Delta Y$ grows. The very first point of both Tables\tref{tab:Y3-2pT0} and\tref{tab:Y5-2pT0}, namely when $|\vec p_H| = |\vec p_J| = 12.5$~GeV, deserves special attention. Here, on one side we are approaching the low-$p_T$ range. On the other side, for $C_\mu = 1/2$, we are very close to the VFNS threshold given by the $b$-quark mass. Therefore the strong dependence on scale variation that we observe both at LLA and NLA indicates that, in this region, our approach has reached its limit of applicability. As a final remark, we notice that our distributions are much smaller when $|\vec p_H| > |\vec p_J|$ than when $|\vec p_H| < |\vec p_J|$. Indeed, it becomes more and more difficult to produce a $b$-flavored bound state than a light jet when the transverse momentum grows.

\begin{table*}[t]
\centering
\caption{Representative values of the double differential $p_T$-distribution [nb/GeV$^2$] for the $H_b$~$+$~jet channel, at $\DY=3$ and $\sqrt{s} = 13$ TeV.}
\label{tab:Y3-2pT0}
\scriptsize
\begin{tabular}{r|r|l|l|l|l|l|l}
  \hline\noalign{\smallskip}
\toprule
$|\vec p_H|$ [GeV] &
$|\vec p_J|$ [GeV] &
$\tarr c \rm LLA \\ C_\mu = 1/2 \earr$ & 
$\tarr c \rm LLA \\ C_\mu = 1 \earr$ &
$\tarr c \rm LLA \\ C_\mu = 2 \earr$ &
$\tarr c \rm NLA \\ C_\mu = 1/2 \earr$ &
$\tarr c \rm NLA \\ C_\mu = 1 \earr$ & 
$\tarr c \rm NLA \\ C_\mu = 2 \earr$ \\
\noalign{\smallskip}\hline\noalign{\smallskip}
\midrule
12.5 & 12.5 &  53.5130(65) & 99.787(28) & 105.835(52) & 56.38(22) & 86.841(96) & 96.14(19) \\
20 & 20 &  9.1732(17) & 10.3279(11) & 9.4236(21) & 9.224(10) & 9.376(15) & 9.147(26) \\
20 & 30 &  4.05134(47) & 4.5122(11) & 4.34781(76) & 3.389(12) & 3.564(12) & 3.630(12) \\
30 & 20 &  2.4020(11) & 2.29978(76) & 1.9232(10) &  
1.373(18) & 1.008(19) & 0.836(19) \\       
30 & 30 &  1.28405(27) & 1.19657(15) & 1.00350(12) & 1.1760(14) & 1.0735(27) & 0.9945(37) \\
30 & 50 &  0.366106(87) & 0.348890(62) & 0.310113(64) & 0.2780(11) & 0.2584(12) & 0.2641(12) \\
50 & 30 &  0.199777(43) & 0.164475(62) & 0.127854(90) & 0.0476(18) & 0.0326(14) & 0.0307(14) \\
50 & 50 &  0.078285(22) & 0.063904(17) & 0.049877(21) & 0.06226(15) & 0.05459(21) & 0.04932(24) \\
75 & 75 &  0.0069352(20) & 0.0052997(16) & 0.0039744(12) & 0.004866(19) & 0.004291(22) & 0.003866(23) \\
\noalign{\smallskip}\hline
\bottomrule
\end{tabular}
\end{table*}
\vskip 0.25cm
\begin{table*}[t]
\centering
\caption{Representative values of the double differential $p_T$-distribution [nb/GeV$^2$] for the $H_b$~$+$~jet channel, at $\DY=5$ and $\sqrt{s} = 13$ TeV.}
\label{tab:Y5-2pT0}
\scriptsize
\begin{tabular}{r|r|l|l|l|l|l|l}
  \hline\noalign{\smallskip}
\toprule
$|\vec p_H|$ [GeV] &
$|\vec p_J|$ [GeV] &
$\tarr c \rm LLA \\ C_\mu = 1/2 \earr$ & 
$\tarr c \rm LLA \\ C_\mu = 1 \earr$ &
$\tarr c \rm LLA \\ C_\mu = 2 \earr$ &
$\tarr c \rm NLA \\ C_\mu = 1/2 \earr$ &
$\tarr c \rm NLA \\ C_\mu = 1 \earr$ & 
$\tarr c \rm NLA \\ C_\mu = 2 \earr$ \\
\noalign{\smallskip}\hline\noalign{\smallskip}
\midrule
12.5 & 12.5 &  22.0870(69) & 32.24231(47) & 29.0534(52) & 10.879(36) & 14.341(17) & 16.446(23) \\
20 & 20 &  2.86919(64) & 2.66456(29) & 2.12591(34) & 1.3263(16) & 1.2985(17) & 1.3380(25) \\
20 & 30 &  0.97469(15) & 0.862500(48) & 0.699142(97) & 0.41688(81) & 0.41494(89) & 0.44790(90) \\
30 & 20 &  0.869094(61) & 0.70087(11) & 0.515064(92) &  0.2345(21) & 0.2010(20) & 0.2036(21) \\       
30 & 30 &  0.314037(38) & 0.249481(27) & 0.186407(24) & 0.132230(16) & 0.12381(23) & 0.12336(29) \\
30 & 50 &  0.0652810(95) & 0.0511772(97) & 0.0390843(50) & 0.023916(84) & 0.023950(83) & 0.025500(73) \\
50 & 30 &  0.056175(16) & 0.0397913(86) & 0.0274650(45) & 0.00676(15) & 0.00681(14) & 0.00814(14) \\
50 & 50 &  0.0135088(22) & 0.0096899(21) & 0.0068649(15) & 0.004823(15) & 0.004661(14) & 0.004635(14) \\
75 & 75 &  0.000843086(72) & 0.000575604(54) & 0.000395652(32) & 0.0002608(14) & 0.0002648(10) & 0.0002661(10) \\
\noalign{\smallskip}\hline
\bottomrule
\end{tabular}
\end{table*}
\vskip 0.25cm

\section{Conclusions and outlook}
\label{sec:conclusions}

We proposed the inclusive emission, in proton-proton collisions, of a forward bottom-flavored hadron accompanied by another backward bottom-flavored hadron or a backward light-flavored jet in semi-hard regimes that can be studied at current LHC energies.

We hunted for signals of stabilization of the high-energy resummation under higher-order corrections and under scale variation, discovering that these effects are present and allow for the description of BFKL-sensitive observables at natural scales, such as the $\DY$-distribution and azimuthal-angle correlations. 
The possibility to study azimuthal moments at natural scales also when jet emissions are allowed is a novel feature which corroborates the statement, already made in the case of $\Lambda_c$ production channels\tcite{Celiberto:2021dzy}, that heavy-flavored emissions of bound states act as fair stabilizers of the high-energy series.
The next part of our program on semi-hard phenomenology relies on a two-fold strategy.

First, we plan to compare observables sensitive to heavy-flavor production in regimes where either the VFNS or the FFNS scheme is relevant, and possibly do a match between the two descriptions. The inclusion of quarkonium production channels, as done in Ref.\tcite{Boussarie:2017oae}, will certainly enrich our phenomenology.

Then, we project an extension of our studies on heavy flavor by considering wider kinematic ranges, as the ones reachable at the EIC\tcite{Accardi:2012qut,AbdulKhalek:2021gbh}, NICA-SPD\tcite{Arbuzov:2020cqg,Abazov:2021hku}, HL-LHC\tcite{Chapon:2020heu}, and the Forward Physics Facility~(FPF)\tcite{Anchordoqui:2021ghd}. Here, the stability of our predictions motivates our interest in (\emph{i}) proposing the hybrid high-energy and collinear factorization as an additional tool to improve the fixed-order description, and (\emph{ii}) evolving our formalism into a \emph{multi-lateral} approach that embody different resummations.

We believe that the study of more and more exclusive observables, such as the double differential transverse-momentum distributions proposed in this work, goes along these directions.

\section*{Acknowledgements}

We thank the Authors of Refs.\tcite{Kniehl:2008zza,Kramer:2018vde} for allowing us to link native {\tt KKSS07} FF routines to the {\tt JETHAD} code\tcite{Celiberto:2020wpk}.
We thank V. Bertone and G. Bozzi for insightful discussions.

F.G.C. acknowledges support from the INFN/NINPHA project and thanks the Universit\`a degli Studi di Pavia for the warm hospitality.
M.F. and A.P. acknowledge support from the INFN/QFT@COLLIDERS project.
The work of D.I. was carried out within the framework of the state contract of the Sobolev Institute of Mathematics (Project No. 0314-2019-0021).

\hypertarget{app:A}{
\section*{Appendix~A: stabilizing effects of $b$-flavor fragmentation}}
\label{app:A}

In this Section we present arguments supporting the statement that $b$-hadron FFs act as stabilizers of our high-energy resummed distributions.

\begin{figure*}[b]

   \includegraphics[scale=0.53,clip]{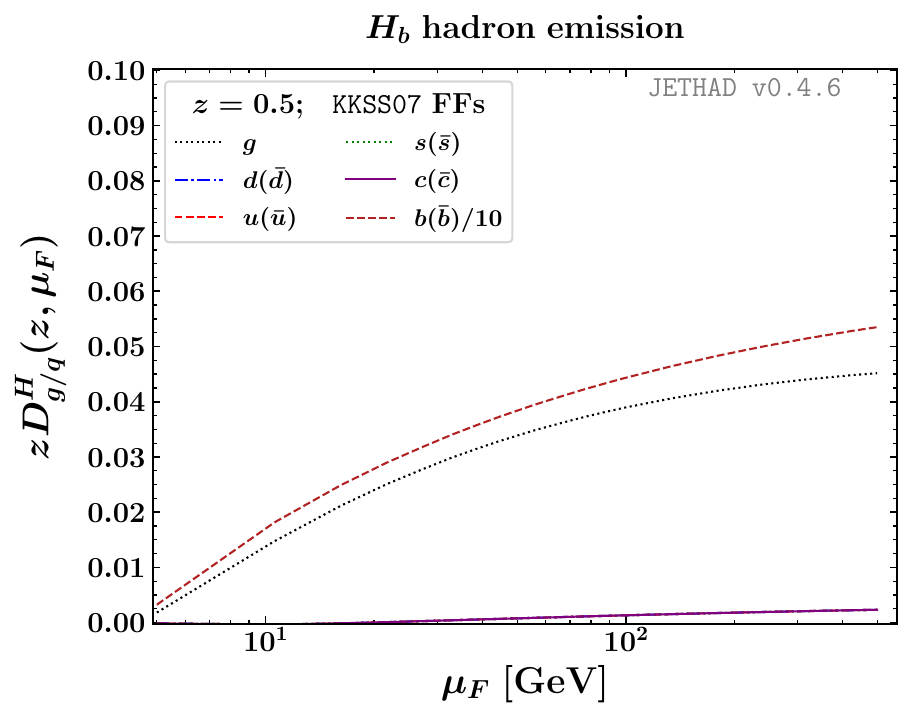}
   \includegraphics[scale=0.53,clip]{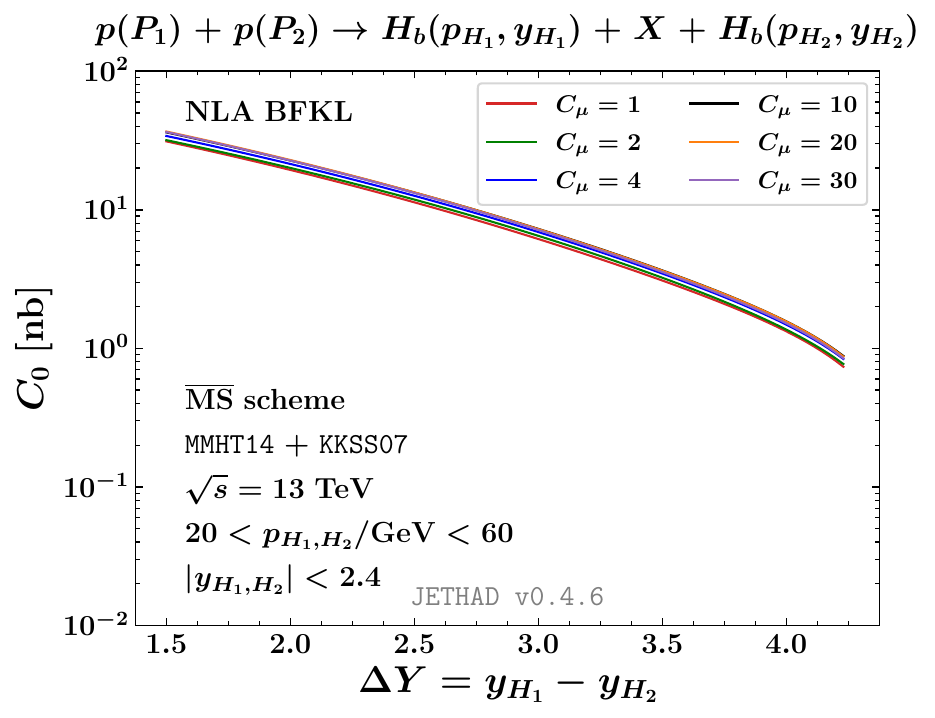}

   \includegraphics[scale=0.53,clip]{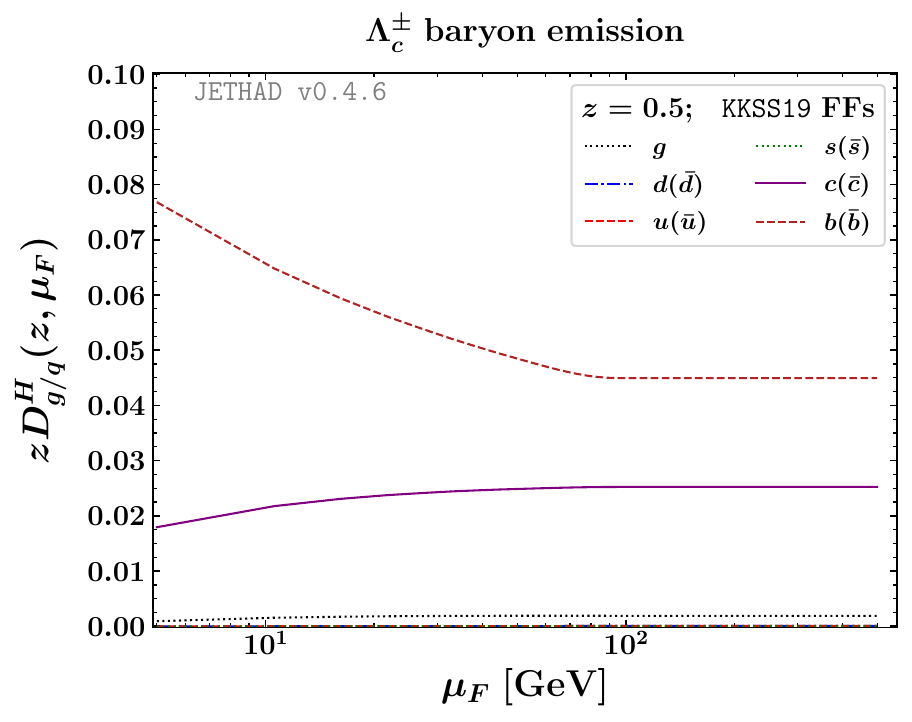}
   \includegraphics[scale=0.53,clip]{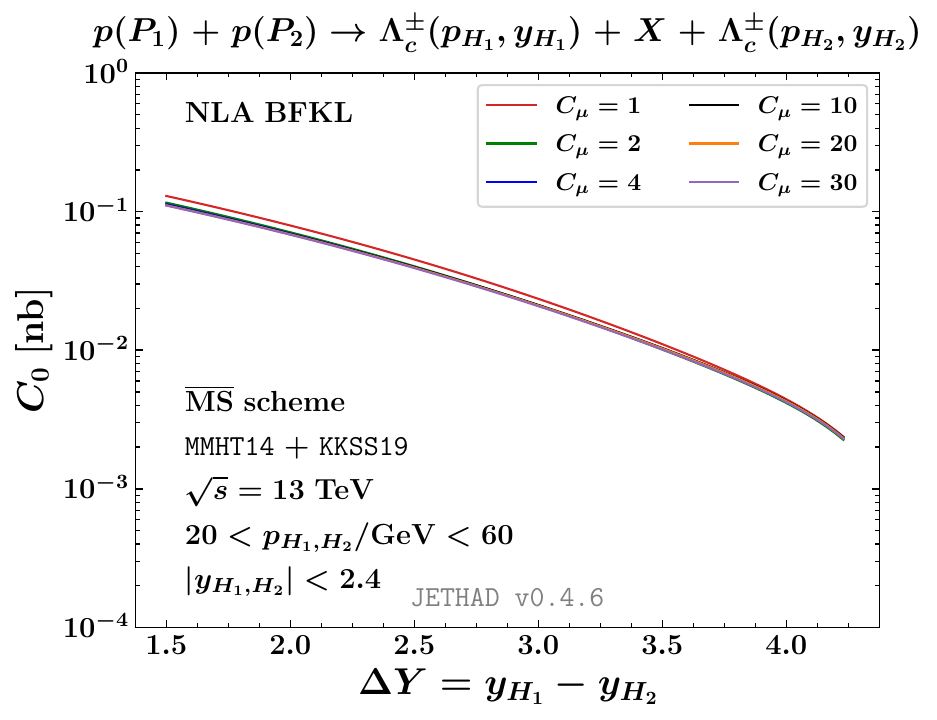}

   \includegraphics[scale=0.53,clip]{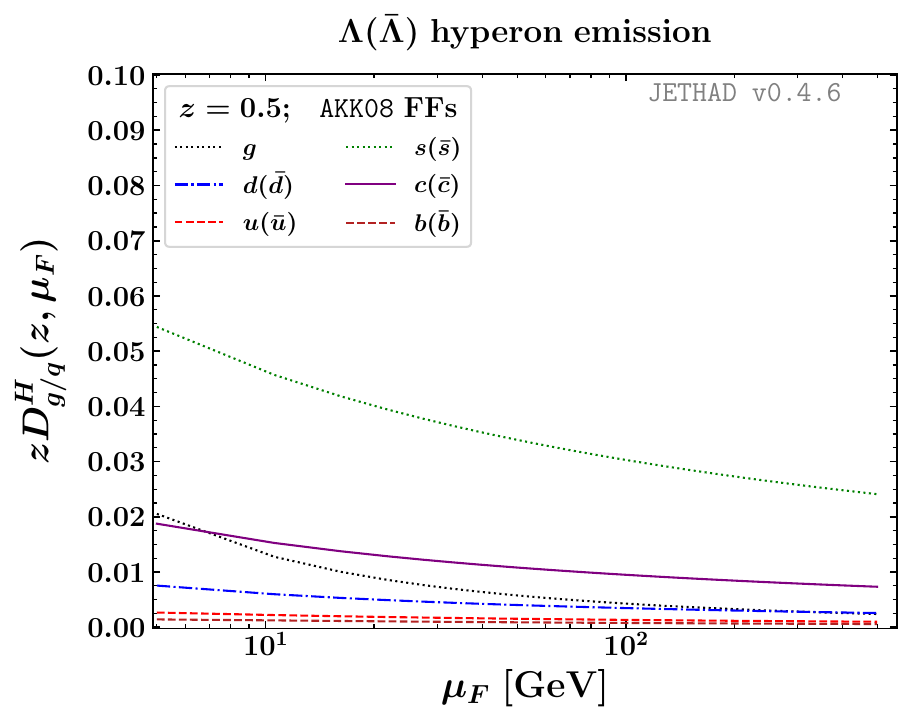}
   \includegraphics[scale=0.53,clip]{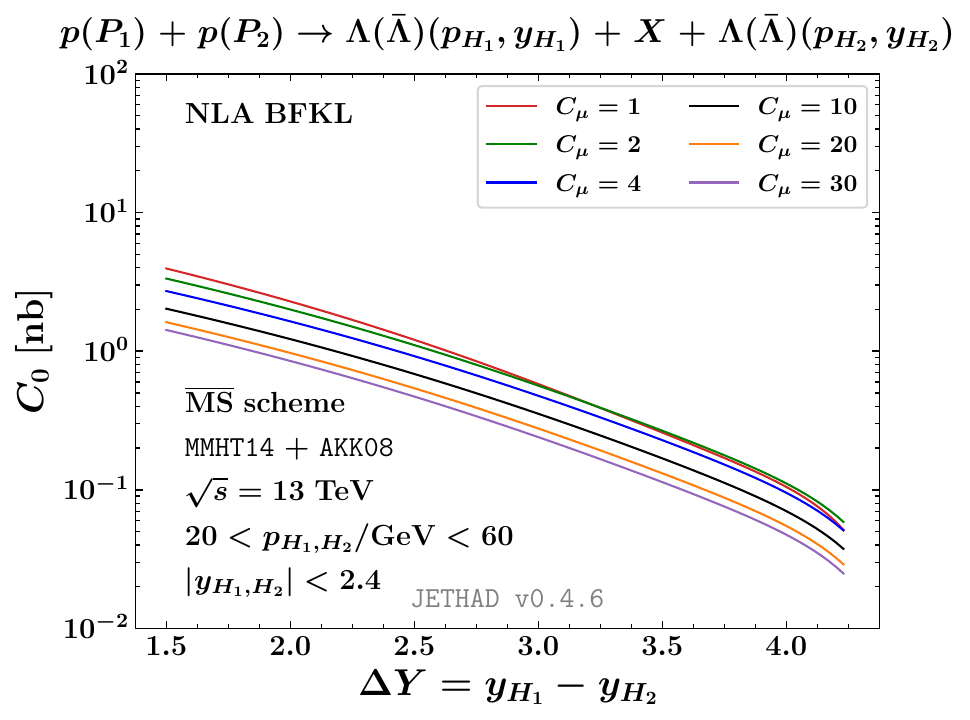}

\caption{Left panels: energy-scale dependence of $H_b$ {\tt KKSS07} (upper, with the $b$-flavor FF reduced by a factor 10), $\Lambda_c^{\pm}$ {\tt KKSS19} (central), and $\Lambda (\bar \Lambda)$ {\tt AKK08} (lower) NLO FFs for $z = 5 \times 10^{-1}$.
Right panels: $\DY$-shape $C_0$ in the double hadron production channel, and for $\sqrt{s} = 13$ TeV. A study on progressive energy-scale variation in the range $1 < C_{\mu} < 30$ is presented for $b$-flavored hadrons (upper), $\Lambda_c$ baryons (central), and $\Lambda(\bar{\Lambda})$ hyperons (lower).}
\label{fig:FFs_C0_psv}
\end{figure*}

In left panels of Fig.\tref{fig:FFs_C0_psv}, going from the top to the bottom, we show the $\mu_F$-behavior of {\tt KKSS07} $H_b$, {\tt KKSS19} $\Lambda_c$, and {\tt AKK08} $\Lambda$ FFs for $z = 0.5 \simeq \langle z \rangle$. This latter roughly corresponds to the average value of $z$ at which FFs are typically probed in kinematic ranges of our analysis. 
We note that the $b$-flavor heavily dominates in $H_b$ fragmentation, the $b$ and $c$-quark ones prevail in $\Lambda_c$ emissions, and the $s$ quark one is on top in $\Lambda$ detections.
However, as pointed out in our previous study on $\Lambda_c$ baryon production (see Section~3.4 of Ref.\tcite{Celiberto:2021dzy}), a dominant role is played by the gluon FF, whose contribution is heightened by the gluon PDF in the diagonal convolution entering LO hadron impact factors~(see Eq.~(\ref{LOIF})) and it holds also at NLO, where $qg$ and $gq$ non-diagonal channels are opened.
In particular, in Ref.\tcite{Celiberto:2021dzy} it was shown how the smooth-behaved, non-decreasing with $\mu_F$ gluon FF depicting the $\Lambda_c$ fragmentation has a strong stabilizing effect on $\DY$-distributions, that also reflects in a partial stabilization of azimuthal-correlation moments.
In this work we confirmed that this feature holds also for $H_b$ hadrons. 

At variance with the $\Lambda_c$ case (Fig.\tref{fig:FFs_C0_psv}, left central panel), the $H_b$ gluon FF clearly grows with $\mu_F$ (Fig.\tref{fig:FFs_C0_psv}, left upper panel). This has an effect on the hierarchy of $C_0$ distributions under a progressive variation of energy scales in the range $1 < C_\mu < 30$, which also includes the typical BLM ones. 
From the inspection of the first two right panels of Fig.\tref{fig:FFs_C0_psv}, it emerges that $C_0$ slightly increases with $C_\mu$ in the double $H_b$ channel, whereas this order is reversed in the $\Lambda_c$ one\footnote{We remark that, in our study on progressive scale variation, expressions for $C_0$ are the ones obtained without applying the BLM prescription, and only the $C_\mu$ factor has been varied.}.
Conversely, the decreasing $\mu_F$-behavior of the gluon {\tt AKK08} function portraying lighter-flavored $\Lambda$ hyperon fragmentation (Fig.\tref{fig:FFs_C0_psv}, left lower panel) leads to an increased sensitivity of results on energy scales (Fig.\tref{fig:FFs_C0_psv}, right lower panel).

All these features corroborate the assertion that the stability of cross sections on energy scales (and then on higher-order corrections), already observed for $c$-flavored emissions\tcite{Celiberto:2021dzy}, is stronger when $b$-flavored bound-state detections are considered. It comes out as the net result of two competing effects. On one hand, higher $\mu_R$ scales (as the BLM ones) make the running coupling smaller, both in the BFKL Green's function and in the impact factors. On the other hand, higher $\mu_F$ values have a mild effect on the $\Lambda_c$ gluon FF, but lead to an increase of the $H_b$ gluon FF. The two features almost compensate each other, and this is the source of the stability that has emerged from our studies.

\begin{figure*}[b]
\centering

   \includegraphics[scale=0.53,clip]{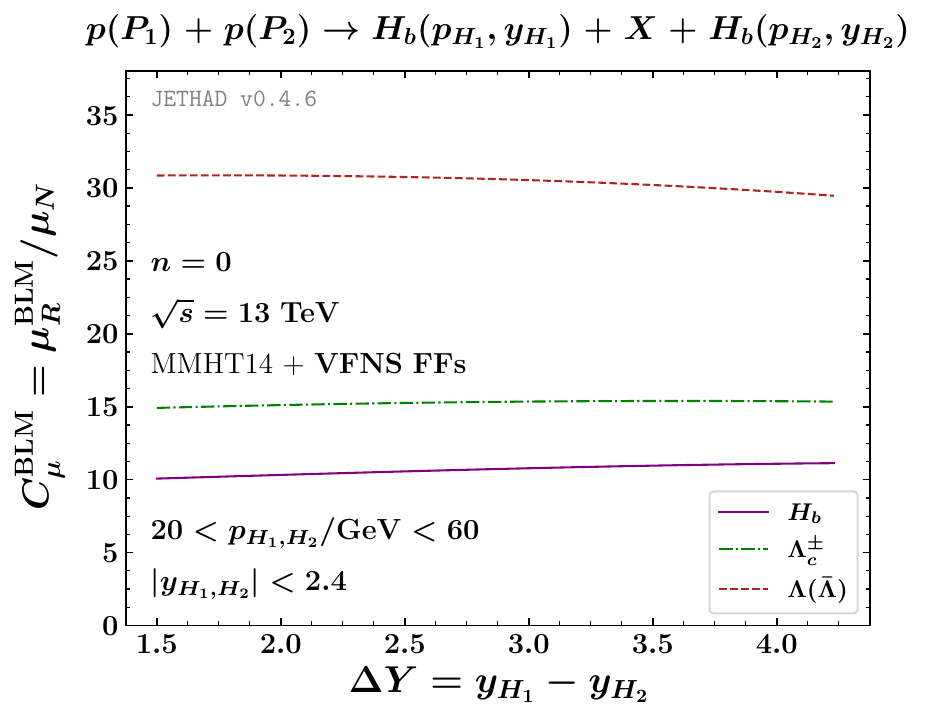}
   \includegraphics[scale=0.53,clip]{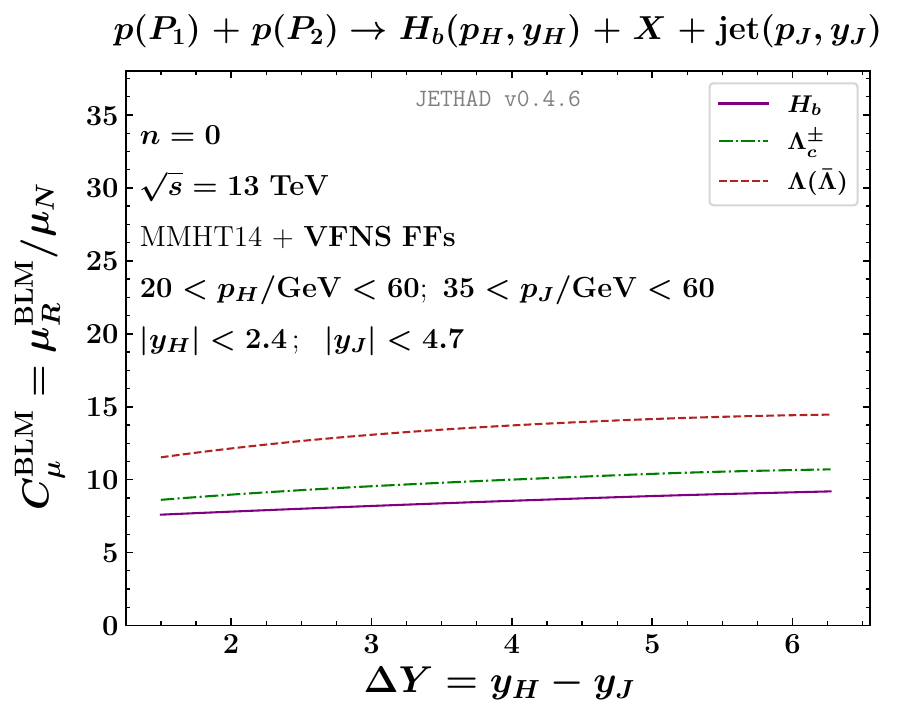}

   \includegraphics[scale=0.53,clip]{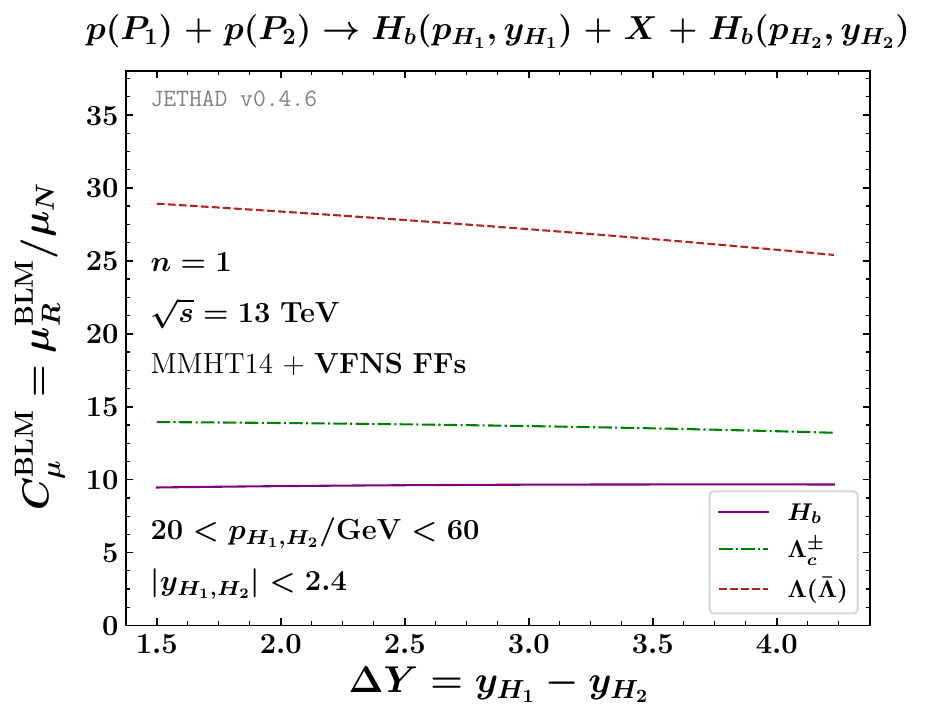}
   \includegraphics[scale=0.53,clip]{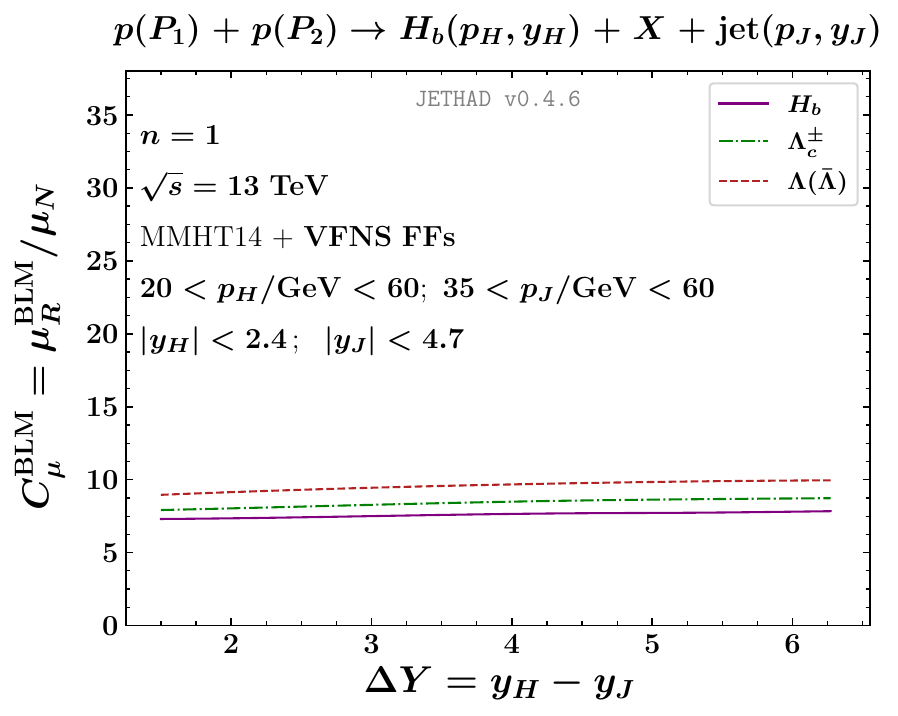}

   \includegraphics[scale=0.53,clip]{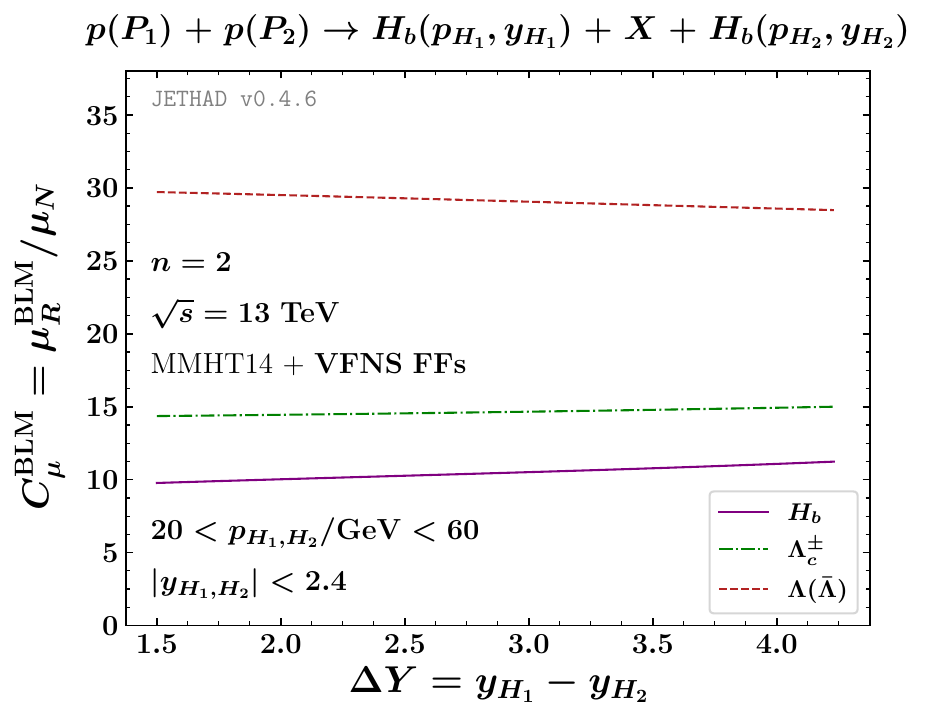}
   \includegraphics[scale=0.53,clip]{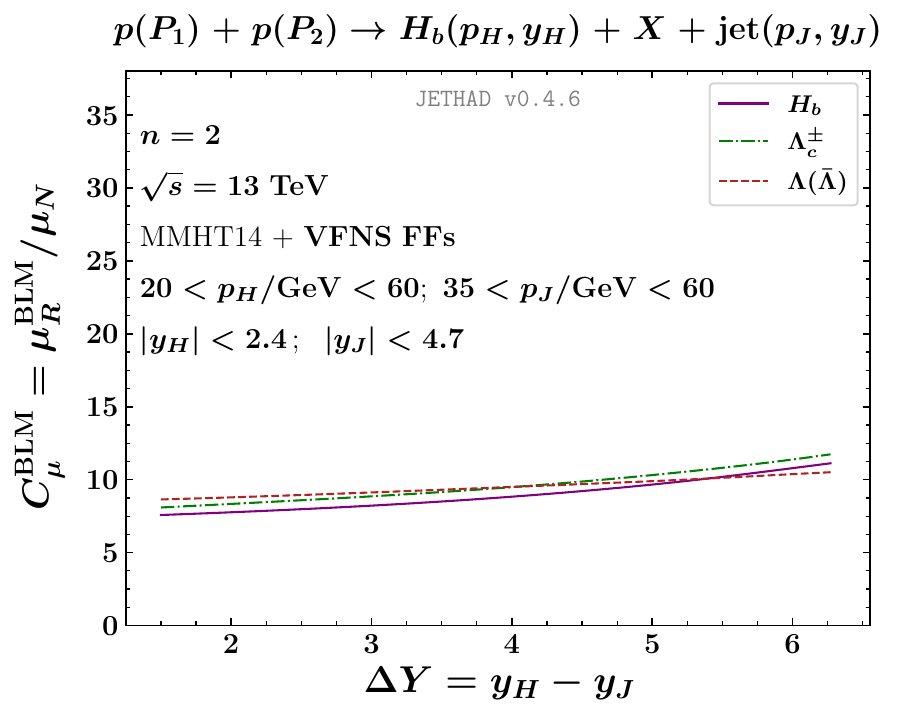}

\caption{BLM scales for the double $H_b$ (left) and the $H_b$~$+$~jet (right) production as functions of the rapidity interval, $\DY$, for $n = 0, 1, 2$, and for $\sqrt{s} = 13$ TeV. Results for $b$-flavor bound states are compared $\Lambda_c$ and $\Lambda(\bar{\Lambda})$ emissions. Text boxes inside panels show transverse-momentum and rapidity ranges.}
\label{fig:BLM_scales_HSA}
\end{figure*}

In Fig.\tref{fig:BLM_scales_HSA} we present the $\DY$-pattern of the BLM-scale parameter, $C_\mu^{\rm BLM}$, for $H_b$ hadrons, $\Lambda_c$ baryons, and $\Lambda$ hyperons. BLM scales for heavy-flavored species are much lower than the ones obtained for lighter hyperons, and the $H_b$ scales are the lowest ones. The effect is much more remarkable in the double hadron channel (left panels) than in the hadron~$+$~jet one (right panels). The found hierarchy between BLM scales is expected. Indeed, since the employment of the BLM scheme operationally translates into a growth of energy-scale values to quench the weight of higher-order corrections, the smaller values of $C_\mu^{\rm BLM}$ are a clue that the high-energy series was already (partially) stable, before adopting BLM.

\bibliography{references}

\end{document}